\documentclass[%
 reprint,
 amsmath,amssymb,
 aps,
 prd,
 superscriptaddress,
]{revtex4-2}

\usepackage{graphicx}% Include figure files
\usepackage{dcolumn}% Align table columns on decimal point
\usepackage{bm}% bold math
\usepackage[%
  colorlinks=true,
  urlcolor=blue,
  linkcolor=blue,
  citecolor=black
]{hyperref}
\usepackage{tikz}
\usetikzlibrary{arrows,shapes,backgrounds,calc,decorations.pathmorphing,patterns,math}
\usepackage[export]{adjustbox}

\newcommand{\sIM}{_{\text{IM}}}
\newcommand{\sCM}{_{\text{CM}}}
\newcommand{\sEM}{_{\text{EM}}}

\newcommand{\sOUT}{_{\text{out}}}
\newcommand{\sopt}{_{\text{opt}}}

\newcommand{\sOM}{_{\text{OM}}}
\newcommand{\sPT}{_{\text{PT}}}

\newcommand{\ssig}{_{\text{sig}}}
\newcommand{\sLO}{_{\text{LO}}}
\newcommand{\XC}{\mathcal{X}}
\newcommand{\ZN}{\mathcal{Z}}
\newcommand{\smin}{_{\text{min}}}
\renewcommand{\Re}{\operatorname{Re}}

\newcommand*\diff{\mathop{}\!\mathrm{d}}

\definecolor{BlockComp}{RGB}{196,210,219}
\definecolor{BlockSpace}{RGB}{255,255,255}
\definecolor{BlockMath}{RGB}{255,255,255}
\definecolor{BlockGain}{RGB}{255,64,64}
\usepackage{dcolumn}
\newcolumntype{d}[1]{D{.}{.}{#1}}

\begin{document}

\title{Enhancing the sensitivity of interferometers with~stable~phase-insensitive~quantum~filters}

\author{Artemiy Dmitriev}
    \email[Corresponding author: ]{admitriev@star.sr.bham.ac.uk}
    \affiliation{
        School of Physics and Astronomy and Institute for Gravitational Wave Astronomy, \\
        University of Birmingham, Edgbaston, Birmingham B15 2TT, United Kingdom
    }
    
\author{Haixing Miao}
    \affiliation{
        School of Physics and Astronomy and Institute for Gravitational Wave Astronomy, \\
        University of Birmingham, Edgbaston, Birmingham B15 2TT, United Kingdom
    }
    \affiliation{Department of Physics, Tsinghua University, Beijing, China 100084}
\author{Denis Martynov}
    \affiliation{
        School of Physics and Astronomy and Institute for Gravitational Wave Astronomy, \\
        University of Birmingham, Edgbaston, Birmingham B15 2TT, United Kingdom
    } 

\begin{abstract}
We present a new quantum control strategy for increasing the shot-noise-limited sensitivity of optical interferometers. The strategy utilizes active phase-insensitive quantum filtering of the signal inside the interferometer and does not rely on optical squeezing. On the example of the coupled-cavity resonators, employed in the gravitational-wave detectors, we show that fully causal and stable phase-insensitive filters can improve the interferometer sensitivity by more than an order of magnitude. The role of the phase-insensitive component in such systems is to provide frequency-dependent compensation for the unwanted dispersion introduced by the position-sensing optical cavity. The system's stability is achieved by limiting the frequency band of this compensation. We demonstrate that stable optomechanical PT-symmetric filters comprise a special subclass of such phase-insensitive devices and find entirely new solutions which overcome the sensitivity of PT-symmetric filters. This scheme is robust against optical loss at the output of the detectors and in the cavities.
\end{abstract}
\maketitle

\section{Introduction}
Quantum nature of light imposes fundamental limits on the sensitivity of laser interferometers~\cite{Clerk2010}, such as axion detectors~\cite{Chaudhuri2018, DeRocco2018, Obata2018, Liu2019, Martynov2020}, optomechanical sensors~\cite{Aspelmeyer2014} and gravitational-wave (GW) observatories~\cite{Fritschel2015, Acernese2014}. In particular, the photon-counting (shot) noise~\cite{Caves1985,Schumaker1985,Kimble2001} leads to the diminished LIGO sensitivity in its most sensitive frequency band above 50\,Hz~\cite{Martynov2016, Buikema2020} and will limit the sensitivity of the proposed third-generation detectors, such as Cosmic Explorer and Einstein Telescope, over their whole sensitivity band~\cite{Reitze2019, ET2020}. New quantum noise suppression techniques can allow us to expand horizons of the axion and GW observatories, improve localization of GW sources~\cite{Klimenko2011}, probe physics of neutron stars~\cite{Martynov2019, Miao2018}, and study black hole spectroscopy~\cite{Berti2016, Laghi2021}.

The GW community widely applies three strategies to suppress the quantum noise: (i) increase the optical power resonating in the detectors, (ii) utilize the coupled-cavity topology~\cite{Meers1988, Mizuno1993}, and (iii) inject squeezed states of light~\cite{Caves1981,Schnabel2017}, or create them internally~\cite{Buonanno2002,Korobko2019}. Each of these techniques has a number of advantages and limitations: (i) leads to optical losses due to the distortion of interferometer geometry caused by the absorption of laser light in the mirrors~\cite{Brooks2021}, creates parametric instabilities~\cite{Braginsky2001,Chen2015,Evans2015}, and increases radiation pressure, which results in the standard quantum limit of sensitivity~\cite{Braginsky1992}, (ii) enhances the detector response above 100\,Hz but suppresses it at lower frequencies (or vice versa)~\cite{Kimble2001,Braginsky2000,Tsang2011}, (iii) improves the quantum noise by making the interferometers quantum-enhanced~\cite{Grote2013,Aasi2013,Acernese2019}, but suffers from optical losses in the detectors~\cite{Kimble2001,Miao2019}.

In this work, we show that another quantum technique known as linear \textit{phase-insensitive} amplification~\cite{Caves1982,Note101} can significantly improve the sensitivity of the optical interferometers. We consider quantum filters given by the equation
\begin{equation}
    \label{eq:ph-ins-amp}
    \hat b = G \hat a + K \hat n_a^\dagger,
\end{equation}
where $\hat{b}$ and $\hat{a}$ are the output and input modes of the filter, $G$ is the gain, and $\hat n_a$ is the filter's internal noise, which is coupled to the output mode with a minimal coupling magnitude of
\begin{equation}
    |K| = \left|\sqrt{\left|G\right|^2-1}\right|.
\end{equation}

\footnotetext[101]{The word ``amplification'' has been a common term for the phase-insensitive processes since the Caves' paper~\cite{Caves1982}. However, Caves' formalism (\ref{eq:ph-ins-amp}) is applicable to systems with any gain $G$, including those with $|G|\le 1$ (e.g. attenuators and filters), which do not ``amplify'' the signal. For most phase-insensitive elements considered in this work, the gain magnitude $|G|$ is either equal to unity or is very close to it; for this reason, to avoid confusion, we use the term ``filters'', rather than ``amplifiers'', when referring to such systems.}

Phase-insensitive amplification has been recently studied in the GW community with the goal to increase the bandwidth of the detectors and to achieve the ``white-light cavity'' effect~\cite{Miao2015,Page2018,Bentley2019,Page2021,Zhang2021,Zhang2021a}. However, these configurations rely on intrinsically unstable systems. In practice, white-light cavities require the development of external stabilization controllers that were not found yet. The next step was made in~\cite{Li2021,Li2021a}, where a stable phase-insensitive amplification based on a PT-symmetric~\cite{Oezdemir2019} optomechanical interaction has been proposed. In this paper, we present causal and stable quantum filters that increase the quantum-limited sensitivity (within a finite frequency band) without compromising the bandwidth or stability of the interferometer. The mechanism underlying this sensitivity enhancement is the frequency-dependent compensation for the unwanted time delay introduced by the sensing cavity. We demonstrate that stable optomechanical PT-symmetric filters found in~\cite{Li2021} form a special sub-class of systems described in this work and find entirely new solutions which overcome the sensitivity limits of PT-symmetric filters. We also consider optical losses at the output of the detector and inside the optical cavities and show that phase-insensitive filters are much more robust against these losses as compared to phase-sensitive techniques, such as the use of squeezed states of light.

\section{Phase-insensitive filtering}
\subsection{Layout}
\begin{figure}%[hbt]
    \centering
    \begin{minipage}[t]{0.03\columnwidth}
        ~a)
    \end{minipage}
    \begin{minipage}[t]{0.95\columnwidth}
        \includegraphics[width=\linewidth,valign=t]{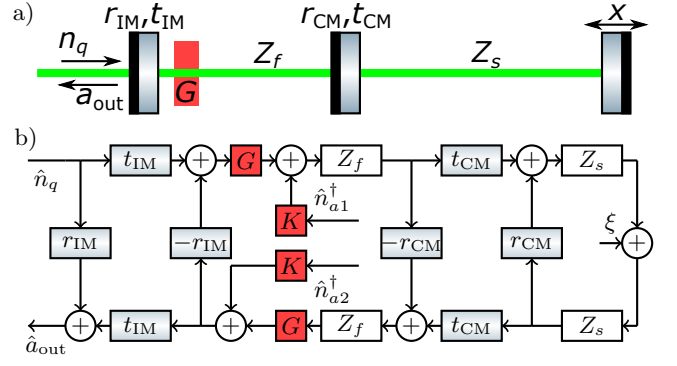}
    \end{minipage}
    \\%[1em]
     \tikzmath{
  \y1 = 0.2; % y of bottom line
  \y2 = 2.4; % y of top line
  \ym = (\y1+\y2) / 2;
 }
 \begin{tikzpicture}[thick]
 \tikzstyle{CompGrad}=[left color = BlockComp,
	right color = BlockComp,
	middle color = white,
	shading angle = 0]

 \draw (0.3,\y2) to node[below]{$\hat n_q$} (0.8,\y2);
 \draw (0.8,\y2) to (1,\y2);
 \draw[->] (1,\y2) to (1.4,\y2);
 \draw [CompGrad] (1.4,\y2-0.2) rectangle node{$t\sIM$} (2.2,\y2+0.2);
 \draw[->] (2.2,\y2) to (2.4,\y2);
 \draw [fill=BlockMath] (2.6,\y2) circle [radius=0.2] node {$+$};
 \draw[->] (2.8,\y2) to (3.0,\y2);
 \draw [fill=BlockGain] (3.0,\y2-0.2) rectangle node{$G$} (3.4,\y2+0.2);
 \draw[->] (3.4,\y2) to (3.6,\y2);
 \draw [fill=BlockMath] (3.8,\y2) circle [radius=0.2] node {$+$};
 \draw[->] (4.0,\y2) to (4.2,\y2);
 \draw [fill=BlockSpace] (4.2,\y2-0.2) rectangle node{$Z_f$} (5.0,\y2+0.2);
 \draw (5.0,\y2) to (5.4,\y2);
 \draw[->] (5.4,\y2) to (5.8,\y2);
 \draw [CompGrad] (5.8,\y2-0.2) rectangle node{$t\sCM$} (6.6,\y2+0.2);
 \draw[->] (6.6,\y2) to (6.8,\y2);
 \draw [fill=BlockMath] (7.0,\y2) circle [radius=0.2] node {$+$};
 \draw[->] (7.2,\y2) to (7.4,\y2);
 \draw [fill=BlockSpace] (7.4,\y2-0.2) rectangle node{$Z_s$} (8.2,\y2+0.2);
 \draw (8.2,\y2) to (8.4,\y2);
 
 \draw[->] (1,\y2) to (1,\ym+0.2);
 \draw [CompGrad] (0.6,\ym-0.2) rectangle node {$r\sIM$} (1.4,\ym+0.2);
 \draw[->] (1,\ym-0.2) to (1,\y1+0.2);
 
 \draw[->] (2.6,\y1) to (2.6,\ym-0.2);
 \draw [CompGrad] (2.2,\ym-0.2) rectangle node {$-r\sIM$} (3.0,\ym+0.2);
 \draw[->] (2.6,\ym+0.2) to (2.6,\y2-0.2);
 
 \draw[->] (3.8,\y2-0.6) to (3.8,\y2-0.2);
 \draw [fill=BlockGain] (3.6,\y2-1.0) rectangle node{$K$} (4.0,\y2-0.6);
 \draw[->] (4.7, \y2-0.8) to node[above] {$\hat n_{a1}^\dagger$} (4.0,\y2-0.8);
 
 \draw[->] (4.7, \y1+0.8) to node[below] {$\hat n_{a2}^\dagger$} (4.0,\y1+0.8);
 \draw [fill=BlockGain] (3.6,\y1+0.6) rectangle node{$K$} (4.0,\y1+1.0);
 \draw (3.6,\y1+0.8) to (3.0,\y1+0.8);
 \draw[->] (3.0,\y1+0.8) to (3.0,\y1+0.2);

 \draw[->] (5.4,\y2) to (5.4,\ym+0.2);
 \draw [CompGrad] (5.0,\ym-0.2) rectangle node {$-r\sCM$} (5.8,\ym+0.2);
 \draw[->] (5.4,\ym-0.2) to (5.4,\y1+0.2);
 
 \draw[->] (7.0,\y1) to (7.0,\ym-0.2);
 \draw [CompGrad] (6.6,\ym-0.2) rectangle node {$r\sCM$} (7.4,\ym+0.2);
 \draw[->] (7.0,\ym+0.2) to (7.0,\y2-0.2);
 
 \draw[->] (8.4,\y2) to (8.4,\ym+0.2);
 \draw [fill=BlockMath] (8.4,\ym) circle [radius=0.2] node {$+$};
 \draw (8.4,\ym-0.2) to (8.4,\y1);
 
 \draw[->] (7.9, \ym) to node[above] {$\xi$} (8.2,\ym);

 \draw[->] (8.4,\y1) to (8.2,\y1);
 \draw [fill=BlockSpace] (7.4,\y1-0.2) rectangle node{$Z_s$} (8.2,\y1+0.2);
 \draw (7.4,\y1) to (7.0,\y1);
 \draw[->] (7.0,\y1) to (6.6,\y1);
 \draw [CompGrad] (5.8,\y1-0.2) rectangle node{$t\sCM$} (6.6,\y1+0.2);
 \draw[->] (5.8, \y1) to (5.6,\y1);
 \draw [fill=BlockMath] (5.4,\y1) circle [radius=0.2] node {$+$};
 \draw[->] (5.2,\y1) to (5.0,\y1);
 \draw [fill=BlockSpace] (4.2,\y1-0.2) rectangle node{$Z_f$} (5.0,\y1+0.2);
 \draw[->] (4.2,\y1) to (4.0,\y1); 
 \draw [fill=BlockGain] (3.6,\y1-0.2) rectangle node{$G$} (4.0,\y1+0.2);
 \draw[->] (3.6,\y1) to (3.2,\y1);
 \draw [fill=BlockMath] (3.0,\y1) circle [radius=0.2] node {$+$};
 \draw (2.8,\y1) to (2.6,\y1);
 \draw[->] (2.6,\y1) to (2.2,\y1);
 \draw [CompGrad] (1.4,\y1-0.2) rectangle node{$t\sIM$} (2.2,\y1+0.2);
 \draw[->] (1.4, \y1) to (1.2,\y1);
 \draw [fill=BlockMath] (1,\y1) circle [radius=0.2] node {$+$};
 \draw[->] (0.8, \y1) to node[below] {$\hat a\sOUT$} (0.3,\y1);
 % Maybe add the carrier?
 \node[anchor=south west] at (0,2.4) {b)};
 \end{tikzpicture}
    \caption{a) General schematic of the detector: a system of two coupled optical resonators, a sensing cavity with length $L_s$ and a filter cavity with length $L_f$. Positional signal $x$ creates a ``signal'' electromagnetic field $\xi$ via modulation of a carrier field, which is resonant in the sensing cavity. The vacuum fluctuations $\hat n_q$ leak into the system through the open port, creating shot noise in the readout mode $\hat a\sOUT$. The active phase-insensitive element $G$ modifies the signal and couples additional noises $\hat n_{a1}^\dagger$ and $\hat n_{a2}^\dagger$ into it. b) Propagation of signal and noise in the system in the Laplace domain.
    \label{fig:qamp_general}}
\end{figure}

We apply phase-insensitive amplification to the coupled cavity layout (Fig.~\ref{fig:qamp_general}a), which is equivalent to the one of the LIGO and Virgo detectors. The layout consists of the input mirror IM, the central mirror CM, and the end mirror EM. Symbols $r$ and $t$ with subscripts IM and CM denote the amplitude reflection and transmission coefficients of the corresponding mirrors, respectively. Motion $x$ of the end mirror (or equivalent GW strain $h=x/L_s$) creates a ``signal'' EM field $\xi$ with a spectral density $S_{\xi\xi}$ via modulation of a carrier field, which is resonant in the sensing cavity with length $L_s$. The filter cavity with length $L_f$ shapes the frequency response of the detector to $\xi$. Information about the signal is read at the output port in the mode $\hat a\sOUT$.

Fig.~\ref{fig:qamp_general}b shows the propagation of the laser fields in the optical system in the Laplace $s$-domain. We propose to implement an active quantum component, characterized by gain $G$ and described by Eq.~(\ref{eq:ph-ins-amp}), in the filter cavity. In our analysis, we assume that the active element $G$ is non-reflective and acts equally on the fields that propagate in both directions. This couples two additional noise sources $\hat n_{a1}^\dagger$, $\hat n_{a2}^\dagger$ into the signal in accordance with (\ref{eq:ph-ins-amp}). The shot noise is accounted for by considering vacuum fluctuations $\hat n_q$ in the input port. In this work, we assume that noise sources $\hat n_{a1}^\dagger$, $\hat n_{a2}^\dagger$, and $\hat n_q$ are all uncorrelated and have equal spectral densities $S_{vv}$. $Z_s(s)=e^{-s\tau_s}$ and $Z_f(s)=e^{-s\tau_f}$ represent phase delay acquired by light as it propagates through the sensing cavity and the filter cavity, respectively. Here $\tau_{s,f} = L_{s,f}/c$ is the one-way propagation time through the respective cavities.

We evolve the signal and noise operators as they propagate through the optical system and compute the transfer functions $T_\xi, T_{\hat n_q}$, $T_{\hat n_{a1}}$,  $T_{\hat n_{a2}}$ from each of these components to the output field, $a\sOUT$ (see Appendix~\ref{app:tf_noloss} for explicit equations of the transfer functions). Our strategy is then to optimize $G$ to achieve the highest signal-to-noise ratio (SNR) in the optical readout. In case of the homodyne readout with the homodyne angle $\phi_{\rm LO}$, which is considered in Appendix~\ref{app:homodyne_readout}, the SNR is given by the equation
\begin{equation}
    \text{SNR}(\omega) = \chi^2(\omega)
    \frac{S_{\xi\xi}(\omega)}{S_{vv}(\omega)},
\end{equation}
where
\begin{equation}
    \chi^2(\omega) = \frac{
        %\left| T_{\hat x} (i\omega) +  T_{\hat x}^* (-i\omega)\right|^2
        \left| 
        T_1 (i\omega) \sin\phi_{\text{LO}}
        +
        i T_2 (i\omega) \cos\phi_{\text{LO}}
        \right|^2
    }
    {
        \sum\limits_{k=\{q,a_1,a_2\}} 
        \left(\left| T_{\hat n_k}(i\omega) \right|^2+
        \left| T_{\hat n_k}(-i\omega) \right|^2\right)
        %\left| T_{\hat n_q}(i\omega) \right|^2+ \left| T_{\hat n_{a1}}(i\omega) \right|^2 + \left| T_{\hat n_{a2}}(i\omega) \right|^2
    }
    \label{eq:chisq}
\end{equation}
and
\begin{equation}
    T_{1,2}(s) = T_{\xi} (s) \pm  T_{\xi}^* (-s).
\end{equation}

\begin{figure}%[hbt]
    \centering
    \includegraphics{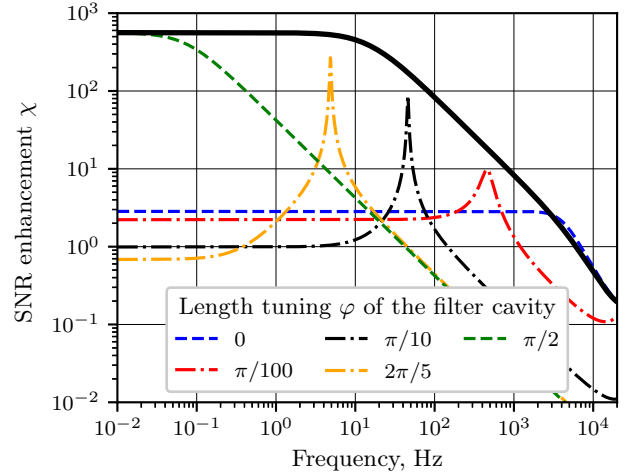}
    \caption{Signal-to-noise ratio enhancement in the optimal readout quadrature of: 1) a passive system ($G\equiv 1$) for different microscopic length tunings $\varphi$ of the filter cavity (thin curves), 2) a system with optimal $G(\omega)$ given by Eq. \ref{eq:Gopt} (thick solid curve).
    \label{fig:passive_and_opt}}
\end{figure}

The subject of our optimization is $\chi^2(\omega)$, which is functionally dependent on $G(i\omega)$. Physically, it represents the SNR enhancement by the interferometer as compared to the system consisting of a single test mirror. 

\subsection{Optimal filter}
We choose the optical homodyne angle $\phi_{\text{LO}}$ and the quantum gain $G$ to achieve the maximum $\chi$. This can be done analytically in the second-order approximation of the delay functions, 
\begin{equation}
    \label{eq:DMA}
    e^{-s\tau}\approx 1- s\tau +s^2\tau^2/2.
\end{equation}
We find (see Appendix~\ref{app:finding_Gopt}) that $\chi^2$ is formally maximal at all frequencies if
\begin{equation}
    \label{eq:Gopt}
    G(s) = G\sopt(s)
    = \sqrt{\frac{s+\gamma_s}{s-\gamma_s}},
\end{equation}
where 
\begin{equation}
    \label{eq:gamma_s}
    \gamma_s = c t\sCM^2/(4L_s)
\end{equation} is the bandwidth of the sensing cavity. This result has an intuitive understanding: the SNR is maximized if $G\sopt^2$ compensates for the unwanted dispersion introduced by the delay $Z_s$ in the sensing cavity~\cite{Zhang2021}, whilst the equation $|G\sopt|\equiv 1$ ensures that the noises $\hat n_{a1}$, $\hat n_{a2}$ do not couple to $\hat a\sOUT$ at all (see Eq.~\ref{eq:ph-ins-amp}). We see that the active phase-insensitive element $G\sopt$ does not act as an ``amplifier'' because it does not change the magnitude of the signal ($|G\sopt|=1$); rather, it comprises an all-pass quantum \textit{filter}. The role of $G\sopt$ is to tune the phase of the laser light in the filter cavity in a frequency-dependent way: $\varphi(\omega)=\arg G\sopt(i\omega)$.

Fig.~\ref{fig:passive_and_opt} shows the comparison of the performance of the optimal filter $G\sopt$ and the passive system ($G\equiv 1$) for the parameters used in~\cite{Li2021a}: $L_s=4$~km, $L_f=40$~m, $T\sEM=0$, $T\sCM=0.5\%$, $T\sIM=2\%$. 
In the passive case, we detune the filter cavity from its resonance by a set of angles in the range from $0$ up to $\pi/2$. The tuning $\varphi=\pi/2$ corresponds to the standard resonant signal recycling technique~\cite{Meers1988}. ``Anti-resonant'' tuning $\varphi=0$ corresponds to the wideband, or ``resonant sideband extraction'' scheme~\cite{Mizuno1993}. Any tuning in between realizes the ``detuned recycling'' scheme~\cite{Buonanno2001}, providing significant enhancement around one particular frequency.  Fig.~\ref{fig:passive_and_opt} shows the enhanced responses in the narrow set of frequencies for different detunings in the passive case. 

In the active case, the gain $G=G\sopt$ ensures optimal response at all frequencies within the sensitive bandwidth (thick solid curve in Fig.~\ref{fig:passive_and_opt}). The response curve has a pole at $s=\gamma_s$. The mechanism of sensitivity enhancement in systems with phase-insensitive amplification is the same as the one provided by passive signal recycling cavities (SRCs), --- namely, compensation for unwanted phase shift generated by the sensing cavity. The difference is that a passive SRC can only provide perfect compensation at a single particular frequency, while in the active phase-insensitive systems such compensation is provided at a wide range of frequencies. Unlike single-sided passive detuned configurations, $G\sopt$ provides optimal phase correction of both signal sidebands. This provides additional enhancement by a factor of $\approx 2$ above the peak values of dash-dotted curves in the figure. Simplified analytic expressions for the sensitivity enhancement are given in Appendix~\ref{app:chi_approx}.

We quantify the total SNR improvement by integrating $\chi^2$ over frequency. In the passive case, the integral's upper bound does not depend on the filter cavity detuning $\varphi$ and can be obtained by using the single-mode approximation:
\begin{equation}
    \int_{0}^{\pi/(2\tau_s)} \chi^2(\omega,G\equiv 1)\diff\omega \lesssim \pi/\tau_s = I_0
\end{equation}
In the active case $G=G\sopt$, the integral enhancement is given by the equation
\begin{equation}
    I\sopt\approx 4\pi /(t\sIM^2\tau_s) = 4 I_0/t\sIM^2
    \label{eq:ILimit}
\end{equation}
and equals $200I_0$ for our set of parameters. 

\paragraph*{Stability.} The optimal gain $G\sopt$ shows a significant SNR enhancement but has an unstable pole at $s = \gamma_s$. In our stability analysis, we use the Nyquist criterion~\cite{Aastrom2021} applied for the open loop transfer function of the whole coupled cavity system shown in Fig.~\ref{fig:qamp_general} (Appendix ~\ref{app:nyquist}).  This analysis shows that the whole system is also unstable if $G=G\sopt$.

However, it is always possible to approximate an unstable $G\sopt$ within a \textit{finite} frequency band with a causal and stable gain $G$ expressed by a rational function with a sufficiently large number of poles and zeros~\cite{Baratchart1997}. If, additionally, the resulting closed-loop transfer function of the whole system is also kept causal and stable, it will still provide sensitivity enhancement in a finite frequency range. We discuss it in the next section.
\section{Stable near-optimal filters}
This section is organized into two parts. First, we show that the coherent quantum feedback scheme with PT-symmetry~\cite{Li2021,Li2021a} is a causal and stable approximation of $G\sopt$ in a finite frequency range. Second, we use a constrained optimization algorithm to find two other stable and causal solutions for $G$ that achieve even stronger SNR enhancement compared to the PT-symmetric systems.

\subsection{PT-symmetric filters}
In the PT-symmetric coherent quantum strategy, the coupled-cavity sensitivity is improved by embedding a mechanical oscillator in the filter cavity with an eigenfrequency $\omega_m$, quality factor $Q_m$, and mass $m$. The filter cavity is then pumped with an addition laser field with frequency $\omega_0 + \omega_m$ and the circulating power $P_f$. The pump field at $\omega_0 + \omega_m$, the mechanical mode at $\omega_m$, and the signal sidebands around $\omega_0$ interact via the radiation pressure. This improves the sensitivity of the detector when the optomechanical coupling rate~\cite{Aspelmeyer2014} is equal to
\begin{equation}
    \label{eq:PT_condition}
    g = \sqrt{\frac{16\pi P_f}{m\lambda\omega_m L_f}} = \frac{\omega_s}{\sqrt{2}},
\end{equation}
where
\begin{equation}
    \omega_s = \frac{t\sCM}{2\tau_f\tau_s}
\end{equation}
is the optical coupling rate between the filter cavity and the sensing cavity.
Moreover, the Hamiltonian of the whole optical system becomes PT-symmetric, allowing it to be stable without any external controllers.

In the resolved sideband regime $\omega_m\gg\gamma_s,\gamma_f$, the PT-symmetric coherent quantum strategy provides phase-insensitive amplification in the filter cavity given by the equation
\begin{equation}
    \label{eq:GOM}
    G(s)=G\sPT(s)=1-\frac{
        4g^2\tau_f\omega_m
    }{
        (s+\gamma_m)(s+\gamma_m-2i\omega_m)
    },
\end{equation}
where $\gamma_m = \omega_m/(2Q_m)$ is the linewidth of the mechanical oscillator. The role of the amplifier's noise is played by the input vacuum noise at frequency $\omega_0 + 2\omega_m$ and the thermal motion of the mechanical oscillator.

We show (see Appendix~\ref{app:opt_OM_coupling} for details) that the PT-symmetric condition (\ref{eq:PT_condition}) corresponds to the case when the gain of the optomechanical component is close to the optimal gain, $G\sPT\approx G\sopt$, for $\omega\gtrsim \gamma_s$. Therefore, at high frequencies $\omega>\gamma_s$, the sensitivity enhancement is limited by
\begin{equation}
    \chi\sPT(\omega)\le \sqrt{2} \frac{t\sCM}{t\sIM}\frac{1}{\omega\tau_s}.
    \label{eq:PTSOMchiHF}
\end{equation}
Details of our analysis of the sensitivity limits in PT-symmetric filters is given in Appendix~\ref{app:PT_limits}. We find that $\arg G\sPT(i\omega)\approx\arg G\sopt(i\omega)$ is satisfied for frequencies above the mechanical linewidth, $\omega\gg\gamma_m$. However, the amplitude requirement  $|G\sPT|\approx|G\sopt|=1$ gives a stronger additional low-frequency limit for the SNR enhancement of the PT-symmetric configuration,
\begin{equation}
    \chi\sPT(\omega) \le \frac{8 t\sIM}{t\sCM^3}\omega\tau_s.
    \label{eq:PTSOMchiLF}
\end{equation}
 Inequalities (\ref{eq:PTSOMchiHF}) and (\ref{eq:PTSOMchiLF}) do not depend on the properties of the mechanical oscillator; in this sense, they reflect fundamental limitations of the PT-symmetric scheme with gain (\ref{eq:GOM}). The dash-dotted curve in Fig.~\ref{fig:3pole} shows a more precise limit of enhancement that can be provided by the PT-symmetric scheme (achieved with $Q_m\to\infty$), which can be approximated by expression (\ref{eq:chi_om_infQ}). The maximal integral enhancement achievable with the PT-symmetric quantum strategy is given by the equation
\begin{equation}
    I_{\text{PT}}
    \approx \pi/\left(t\sIM\tau_s\right)
    =I_0/t\sIM
\end{equation}
and equals to $\approx 7$ for our set of parameters.

The PT-symmetric optomechanical element $G\sPT$ is causal and stable since its gain~(\ref{eq:GOM}) has no poles in the right-hand side complex half-plane. Stability analysis of the whole system is somewhat complicated because of the ``idler'' mode $\omega_0+2\omega_m$, which also circulates in the interferometer and must be included in the stability analysis; such rigorous analysis was presented in~\cite{Li2021a}, where it was shown that the full system is also causal and stable.

However, we show below that the approximation of $G\sopt$ achieved by the optomechanical PT-symmetric filters is not optimal, even among causal and stable approximations of $G\sopt$ with rational functions that have only two poles.

\subsection{Constrained optimisation of sensitivity}
\label{sec:constrained_opt}
We developed an algorithm to find optimal zeros $Z$, poles $P$, and gain $K$ of the approximation of $G\sopt$ to maximize the integral given by the equation
\begin{equation}
    I_{\chi^2} = \frac{1}{I_0}\int_{0}^{\pi/\tau_s} \chi^2(\omega,Z,P,K)\diff\omega.
\end{equation}
The optimization algorithm also enforces the stability of the element $G$ and the stability of the coupled-cavity system. This is achieved by detecting open-loop instability (i.e. unstable poles of Eq.~\ref{eq:TOL}) and closed-loop instability (using an automated Nyquist stability criterion) and including them in the cost function with large weights. A detailed description of our algorithm and the solutions we obtained with it are given in Appendix~\ref{app:optimization}.

\begin{figure}[t]%[hbt]
    \centering
    \includegraphics{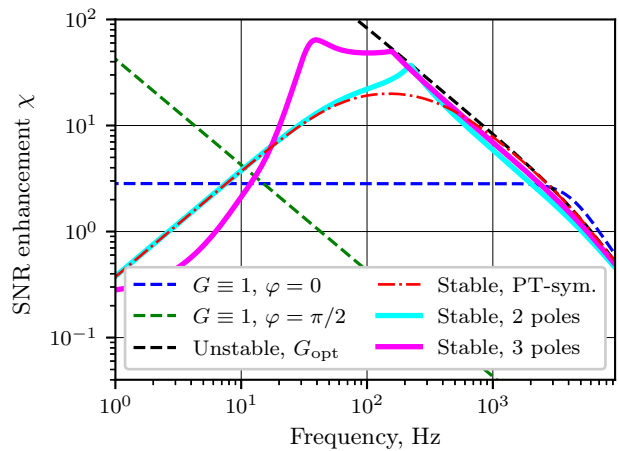}
    \caption{Thick lines represent stable optimized configurations of $G$ with 2 poles (cyan) and 3 poles (magenta) with improved integral enhancement $I_{\chi^2}$ as compared to the best PT-symmetric OM configuration (dash-dotted line).
    \label{fig:3pole}}
\end{figure}

\paragraph{Two-pole systems.} We first searched for the optimal $G$ with two poles and utilized the transfer function given by Eq.~(\ref{eq:GOM}) as an initial guess. Using our constrained optimization algorithm, we found a stable phase-insensitive filter with $I_{\chi^2}\approx 9.2$\,dB, which exceeds the PT-symmetric limit of sensitivity enhancement $I_{\text{PT}}\approx 8.5$\,dB (cyan curve in Fig.~\ref{fig:3pole}).

\paragraph{Three-pole systems.}We further improved SNR enhancement, $\chi$, by increasing the number of poles of the phase-insensitive element. The magenta curve in Fig.~\ref{fig:3pole} shows $\chi$ for the case of a $G$ with three stable poles. The pole and zero locations were obtained using the \texttt{vectfit} algorithm~\cite{Gustavsen1999} modified for complex-valued impulse responses and further improved using the constrained optimization described above, resulting in $I_{\chi^2}\approx 12.5$\,dB.

\section{Effect of optical losses}
The key advantage of our quantum noise reduction strategy is its resilience to optical losses. To demonstrate it, we employ the phase-insensitive formalism: for each optical loss channel with loss $\Lambda$ we add a quantum phase-insensitive component described by (\ref{eq:ph-ins-amp}) with gain $G=\sqrt{1-\Lambda}$. Specifically, in this work, we consider the output loss $\Lambda_o$, the round-trip filter cavity loss $\Lambda_f$, and the round-trip sensing cavity loss $\Lambda_s$. Each of these loss channels couples in an additional quantum noise process $\hat n_\Lambda^\dagger$; we assume that these noise processes are uncorrelated between each other and can be described with the same spectral density (that of vacuum fluctuations) $S_{vv}$. These noise sources, together with previously considered shot noise $\hat n_q$ and noise fluctuations associated with the phase-insensitive element,$\hat n_{a1}^\dagger$ and $\hat n_{a2}^\dagger$, all contribute to the denominator of the sensitivity enhancement figure~(\ref{eq:chisq}) through their corresponding transfer functions, which are given and discussed in Appendix~\ref{app:tf_loss}. 

\begin{figure}[hbt]
    \centering
    \includegraphics{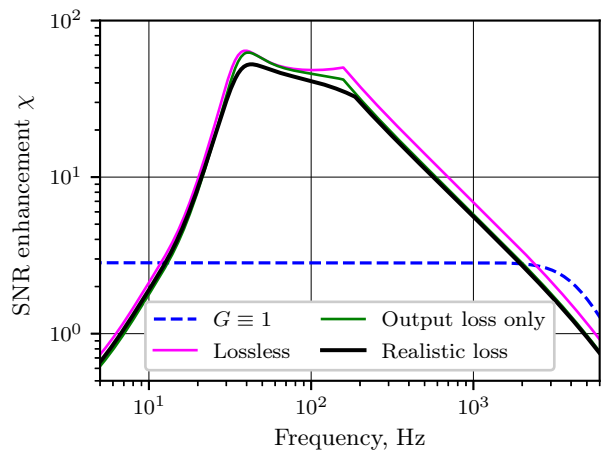}
    \caption{Comparison of the SNR enhancement $\chi$ provided by stable optimized configurations of $G$ with 3 poles with and without optical losses. An output loss of $\Lambda_o=30\%$ (green curve) reduces the integral SNR $I_{\chi^2}$ from the lossless case (magenta curve) by $20\%$. A combination of $\Lambda_o=30\%$, filter cavity loss $\Lambda_f=0.2\%$, and sensing cavity loss $\Lambda_s=0.005\%$ (thick black curve) decreases $I_{\chi^2}$ by $34\%$ from the lossless case. All configurations retain both open-loop and closed-loop stability.
    \label{fig:lossy}}
\end{figure}

Optical loss is the main limiting factor for the performance of phase-sensitive techniques (optical squeezing) that are currently in use in GW detectors. In particular, optical loss $\Lambda_o$ in the output channel decreases the measurable level of squeezing from infinity to $\lesssim 6$\,dB for $\Lambda_o=30\%$~\cite{Dwyer2013}. However, the same amount of output loss will reduce the SNR enhancement $\chi^2$ only marginally in a system based on a phase-insensitive element in the filter cavity. This is illustrated by Fig.~\ref{fig:lossy}, where the integral enhancement $I_{\chi^2}$ is decreased by $\approx 20\%$ by output loss (green solid curve). Furthermore, the thick black curve in Fig.~\ref{fig:lossy} shows the enhancement provided by a stable configuration with a 3-pole transfer function $G$ of the filter and optical losses typical for a modern GW detector: $\Lambda_o=30\%$, $\Lambda_f=0.2\%$~\footnote{The actual optical loss in the filter cavity will depend on the physical realization of $G$, which is beyond the scope of this paper. Here we choose $\Lambda_f=0.2\%$ as a reference value, to show that neither SNR enhancement nor the stability of the system is compromised in the presence of such an effective loss in the filter cavity.}, and $\Lambda_s=50$\,ppm. The integral enhancement $I_{\chi^2}$ is decreased by $\approx34\%$ with respect to the lossless case, but it still exceeds the enhancement of schemes without phase-insensitive amplification ($G\equiv 1$) by a factor of $\approx 12$.

\section{Conclusion}
We have shown that stable quantum phase-insensitive filters can significantly improve the SNR of coupled-cavity systems. Such filters can be implemented by means of optomechanical resonators discussed above, or via quantum conversion of optical frequency in nonlinear crystals~\cite{Vollmer2014}. The system configuration can be obtained from a known $G$ via the quantum network synthesis method~\cite{Bentley2021}; however, detailed studies of the technical design lie beyond the scope of the present paper. The use of quantum photonic integrated circuits~\cite{Dietrich2016,Elshaari2020} can allow for stable broadband systems with many optical poles, whose integral SNR enhancement approaches the limit (\ref{eq:ILimit}).

In contrast to the injection of squeezed states of light, which improves the SNR by reducing the variance of the vacuum noise in the measured quadrature, stable phase-insensitive amplification improves the SNR by enhancing the signal field in the interferometer. Therefore, the two quantum techniques are complementary and can be used together to improve the quantum noise-limited sensitivity of interferometric detectors. The sensitivity can be improved even further by introducing quantum correlations between different noise sources (in particular, the shot noise $\hat n_q$ and the noise of the active element $\hat n_{a1,2}$. We leave these topics for future research.

\begin{acknowledgments}
We thank George Smetana, Teng Zhang, Amit Ubhi, Joe Bentley, Chunnong Zhao for their support, and members of the LSC Quantum Noise and Advanced Interferometer Configuration working groups for fruitful discussions. We acknowledge the support of the Institute for Gravitational Wave Astronomy at the University of Birmingham, STFC Quantum Technology for Fundamental Physics scheme (Grant No. ST/T006609/1), and EPSRC New Horizon Scheme (Grant No. EP/V048872/1). H.M. is supported by UK STFC Ernest  Rutherford Fellowship (Grant No. ST/M005844/11).
\end{acknowledgments}

\appendix
\section{Transfer functions}
\label{app:tf}
\subsection{The lossless case}
\label{app:tf_noloss}
In this section, we obtain explicit expressions for the transfer functions to the output field $a\sOUT$~(Fig.~\ref{fig:qamp_general}): $T_\xi$ from the signal $\xi$ and $T_{\hat n_q}$, $T_{\hat n_{a1}}$, and $T_{\hat n_{a2}}$ from each of the sources $\hat n_{a1}^\dagger$, $\hat n_{a2}^\dagger$, and $\hat n_q$ of the quantum noise. The arm cavity is assumed to be resonant with the carrier light with frequency $\omega_0$. Since there is an active element with a complex-valued gain $G$ in the filter cavity, we assume without loss of generality that the filter cavity without $G$ would also be resonant at $\omega_0$ because we include any additional phase shift in the filter cavity into the expression for $G$. This corresponds to the ``resonant sideband extraction'' tuning used in Advanced LIGO and results in an increased bandwidth at the cost of limited peak sensitivity (see the case $\varphi=0$ in Fig.~\ref{fig:passive_and_opt}).

The key element of this setup is an active phase-insensitive component for the optical mode embedded into the filter cavity. The component's gain $G(s)$ is represented by real-valued gain magnitude $G_0$ and phase $\varphi_G$ as
\begin{equation}
    G(s) = G_0(s) e^{i \varphi_G (s)}.
\end{equation}
In accordance with Caves' formalism, such a phase-insensitive filter introduces additional noise $\hat n_a^\dagger$ whose impact has a lower bound which depends on the gain magnitude, and the relation between the input mode $\hat a$ and the output mode $\hat b$ is
\begin{equation}
    \hat b=G\hat a+\left|\sqrt{G_0^2-1}\right|\hat n_a^\dagger.
    \label{eq:ph-ins-amp-app}
\end{equation}
Quantum mechanics imposes a constraint only on the magnitude of the internal mode's coupling to the output. The phase term can be arbitrary, and we ignore it as irrelevant for now.

In this document, the following convention for the Fourier transform is assumed:
\begin{equation}
    \tilde x(\omega)=F[x(t)](\omega) = \int\limits_{-\infty}^{\infty}x(t)e^{-i\omega t}\diff t.
\end{equation}

The four transfer functions, one from from the signal $\hat\xi$ and three from the three noises $\hat n_q$,$\hat n_{a1}$,$\hat n_{a2}$ to the output $\hat a\sOUT$, are
\begin{widetext}
\begin{align}
    T_\xi = \left(\frac{\hat a\sOUT}{\hat \xi}\right)_{\hat n_q, \hat n_{a1}, \hat n{a_2} \equiv 0} &=  
    -\frac{G_0 e^{i\varphi_G} t\sCM t\sIM Z_f Z_s}
    {-1+r_{CM}Z_s^2 + G_0^2e^{2i\varphi_G}r\sIM Z_f^2(r\sCM-Z_s^2)},\\ 
    T_{n_q} = \left(\frac{\hat a\sOUT}{\hat n_q}\right)_{\hat \xi, \hat n_{a1}, \hat n{a_2} \equiv 0} &=  
    \frac{G_0^2 e^{2i\varphi_G} Z_f^2 (r\sCM - Z_s^2) + r\sIM(-1+r\sCM Z_s^2)}
    {-1+r_{CM}Z_s^2 + G_0^2e^{2i\varphi_G}r\sIM Z_f^2(r\sCM-Z_s^2)},\\ 
    T_{n_{a1}} = \left(\frac{\hat a\sOUT}{\hat n_{a1}}\right)_{\hat \xi, \hat n_q, \hat n{a_2} \equiv 0} &=  
    \frac{e^{i\varphi_G}G_0\sqrt{1-G_0^2} t\sIM Z_f^2 (r\sCM - Z_s^2)}
    {-1+r_{CM}Z_s^2 + G_0^2e^{2i\varphi_G}r\sIM Z_f^2(r\sCM-Z_s^2)}, \\
    T_{n_{a2}} = \left(\frac{\hat a\sOUT}{\hat n_{a2}}\right)_{\hat \xi, \hat n_q, \hat n{a_1} \equiv 0} &=  
    \frac{\sqrt{1-G_0^2} t\sIM (-1+ r\sCM Z_s^2)}
    {-1+r_{CM}Z_s^2 + G_0^2e^{2i\varphi_G}r\sIM Z_f^2(r\sCM-Z_s^2)}.
\end{align}
\end{widetext}

\subsection{Effect of optical losses}
\label{app:tf_loss}
In this work, we consider three types of optical losses:
\begin{itemize}
    \item Output loss $\Lambda_o$,
    \item Filter cavity loss $\Lambda_f$,
    \item Sensing cavity loss $\Lambda_s$.
\end{itemize}
From the quantum mechanical point of view, any loss $\Lambda$ is also a phase-insensitive component with gain $\sqrt{1-\Lambda}$ and additional quantum noise $\hat n_\Lambda$, which gets coupled to the output signal $\hat b$ through coefficient $\sqrt{\Lambda}$:
\begin{equation}
    \hat b = \sqrt{1-\Lambda}\hat a + \sqrt{\Lambda}\hat n_\Lambda^\dagger,
\end{equation}
where $\hat a$ is the input mode (cf. Eq. \ref{eq:ph-ins-amp}).

In order to calculate the transfer functions to the output, we amend the configuration in Fig.~\ref{fig:qamp_general} by adding loss elements $\sqrt{1-\Lambda_o}$, $\sqrt{1-\Lambda_f}$, and $\sqrt{1-\Lambda_s}$ and their corresponding noises $\hat n_{\Lambda_o}^\dagger$, $\hat n_{\Lambda_f}^\dagger$, $\hat n_{\Lambda_s}^\dagger$ to the output signal, the filter cavity, and the sensing cavity, respectively. Introducing the modified propagators $\ZN_f=Z_f \sqrt{1-\Lambda_f}$, $\ZN_s = Z_s \sqrt{1-\Lambda_s}$ and the output effectiveness $\eta = \sqrt{1-\Lambda_o}$, we obtain
\begin{widetext}
\begin{align}
    T_\xi &=  
    -\frac{\eta G_0 e^{i\varphi_G} t\sCM t\sIM \ZN_f \ZN_s}
    {-1+r_{CM}\ZN_s^2 + G_0^2e^{2i\varphi_G}r\sIM \ZN_f^2(r\sCM-\ZN_s^2)},\\ 
    T_{n_q} &=  
    \frac{\eta\left[G_0^2 e^{2i\varphi_G} \ZN_f^2 (r\sCM - \ZN_s^2) + r\sIM(-1+r\sCM \ZN_s^2)\right]}
    {-1+r_{CM}\ZN_s^2 + G_0^2e^{2i\varphi_G}r\sIM \ZN_f^2(r\sCM-\ZN_s^2)},\\ 
    T_{n_{a1}} &=  
    \frac{\eta e^{i\varphi_G}G_0\sqrt{1-G_0^2} t\sIM \ZN_f^2 (r\sCM - \ZN_s^2)}
    {-1+r_{CM}\ZN_s^2 + G_0^2e^{2i\varphi_G}r\sIM \ZN_f^2(r\sCM-\ZN_s^2)},\\
    T_{n_{a2}} &=  
    \frac{\eta \sqrt{1-\Lambda_f} \sqrt{1-G_0^2} t\sIM (-1+ r\sCM \ZN_s^2)}
    {-1+r_{CM}\ZN_s^2 + G_0^2e^{2i\varphi_G}r\sIM \ZN_f^2(r\sCM-\ZN_s^2)},\\
    T_{n_{\Lambda o}} &= \sqrt{\Lambda_o},\\
    T_{n_{\Lambda f}} &= \frac{\eta \sqrt{\Lambda_f} t\sIM (-1+ r\sCM \ZN_s^2)}
    {-1+r_{CM}\ZN_s^2 + G_0^2e^{2i\varphi_G}r\sIM \ZN_f^2(r\sCM-\ZN_s^2)},\\
    T_{n_{\Lambda s}} &= -\frac
    {\eta G_0 e^{i\varphi_G} t\sCM t\sIM \ZN_f Z_s \sqrt{\Lambda_s}}
    {-1+r_{CM}\ZN_s^2 + G_0^2e^{2i\varphi_G}r\sIM \ZN_f^2(r\sCM-\ZN_s^2)}.
\end{align}
\end{widetext}
To calculate the SNR enhancement, we include terms associated with noises $\hat n_{\Lambda_o}^\dagger$, $\hat n_{\Lambda_f}^\dagger$, $\hat n_{\Lambda_s}^\dagger$ introduced by optical losses into (\ref{eq:PSD_noise}).

The presence of output loss $\Lambda_o$ does not affect the internal stability of the system. On the other hand, filter cavity loss $\Lambda_f$ and sensing cavity loss $\Lambda_s$ do affect the stability. Therefore, we need to apply the optimization algorithm described in Appendix~\ref{app:optimization_algorithm} separately to each particular set of loss coefficients. Results for $\Lambda_f=0.2\%$, $\Lambda_s=50$\,ppm are presented in Table~\ref{tab:zpk}.

\section{SNR enhancement of the homodyne readout}
\label{app:homodyne_readout}
\subsection{Signal}
\paragraph{Modulation.} In an interferometer, signal $\xi$ is created by phase modulation of carrier field with amplitude $A_c$ and frequency $\omega_0$ by the effective motion $x$ of a test mirror. For simplicity, in our derivation we will consider classical mechanical motion at a single frequency $\omega$ only:
\begin{equation}
    x(t) = x_\omega \cos(\omega t+\phi_\omega) = \frac{1}{2}
    \left(
        X_\omega e^{i\omega t} + X_\omega^* e^{-i \omega t}
    \right),
\end{equation}
where $X_\omega = x_\omega e^{i \phi_\omega}$.

The field reflected of the test mirror, 
\begin{equation}
    a_{\text{reflected}}(t) = a_{\text{incident}}(t-\frac{2x(t)}{c})\approx a(t) - \frac{2x(t)}{c} \frac{\partial a}{\partial t},
\end{equation}
 contains modulation sidebands, which comprise the signal $\xi(t)$:
\begin{multline}
    \xi(t) = -\frac{2\pi i}{\lambda_0}A_c
    \left(
        X_\omega e^{i(\omega_0+\omega) t} +
        X_\omega^* e^{i(\omega_0-\omega)t}
    \right) + c.c. \\ = 
    -i\left(
        \xi_\omega e^{i(\omega_0+\omega) t} +
        \xi_\omega^* e^{i(\omega_0-\omega)t}
    \right) +c.c.,
\end{multline}
where $\lambda_0 = \omega_0/c$ and the symbol $c.c.$ here and below denotes the complex conjugate of the preceding expression.

\paragraph{Output.} The quantity $\xi(t)$ serves as a classic input to the interferometer. By switching to the frequency domain, propagating this signal to the output $a\sOUT$ of the interferometer, and switching back to the time domain, we find the signal component $a\ssig$ of the output:
\begin{multline}
    a\ssig(t)=\sin(\omega_0 t)
    \left[
        \chi_1 e^{i\omega t} + \chi_1^* e^{-i\omega t}
    \right] \\
    + \cos(\omega_0 t)
    \left[
        \chi_2 e^{i\omega t} + \chi_2^* e^{-i\omega t}
    \right] \\
    = 
    \alpha\ssig (t) \sin(\omega_0 t) + \beta\ssig (t) \cos(\omega_0 t).
    \label{eq:aout_sig}
\end{multline}
Here
\begin{align}
    \chi_1(\omega) &= \frac{\chi_+(\omega)+\chi_-(\omega)}{\sqrt 2},\\
    \chi_2(\omega) &= \frac{\chi_+(\omega)-\chi_-(\omega)}{i\sqrt 2}
\end{align}
are the ``phase'' and the ``amplitude'' field quadratures, respectively, and
\begin{align}
    \chi_+(\omega) &= \sqrt{2} \xi_\omega T_\xi(i\omega),\\
    \chi_-(\omega) &= \sqrt{2} \xi_\omega^* T_\xi(-i\omega).
\end{align}

\paragraph{Readout.} In the balanced homodyne setup, the output field $a\ssig$ (\ref{eq:aout_sig}) beats with a local oscillator field
\begin{equation}
    a\sLO = \varepsilon \cos{(\omega_0 t +\phi\sLO)},
    \label{eq:aLO}
\end{equation}
where $\phi\sLO$ is the homodyne angle. The two fields beat with each other on a 50:50 beamsplitter, creating another pair of fields with amplitudes proportional to $a\sOUT\pm a\sLO$. These two fields are incident on two photodetectors, creating photocurrents $i_1$ and $i_2$. The currents are proportional to the average power of the incident fields per oscillation:
\begin{equation}
    i_{1,2}(t) \propto \overline{(a\sOUT\pm a\sLO)^2} = 
    \frac{\omega_0}{2\pi}
    \int_0^{2\pi/\omega_0}(a\sOUT\pm a\sLO)^2 \diff t.
    \label{eq:photocurrents}
\end{equation}
Substituting (\ref{eq:aout_sig}) and (\ref{eq:aLO}) into (\ref{eq:photocurrents}) and taking the difference between the two photocurrents, we obtain the readout signal
\begin{equation}
    i_\text{signal} = i_1 - i_2 \propto
    \alpha(t) \cos{\phi\sLO} - \beta(t) \sin{\phi\sLO}.
\end{equation}
Calculating the power spectral density (PSD) $S\ssig$ of this signal, we get
\begin{multline}
    \frac{S\ssig(\omega)}{S_{\xi\xi}(\omega)} \propto \big|
        \left(
            T_\xi(i\omega)+T_\xi^*(-i\omega)
        \right) \sin{\phi\sLO} 
        \\ + 
        i \left(
            T_\xi(i\omega)-T_\xi^*(-i\omega)
        \right) \cos{\phi\sLO}
    \big|^2,
    \label{eq:PSD_signal}
\end{multline}
which constitutes the numerator of the right hand side of Equation~\ref{eq:chisq}.
\subsection{Noise}
In order to calculate the noise, we apply standard two-photon formalism suggested in~\cite{Caves1985,Schumaker1985} and described in detail for coupled-cavity systems in~\cite{Kimble2001}; the only difference from the latter paper is that we have three separate incoherent noise sources ($\hat n_q$, $\hat n_{a1}$, $\hat n_{a2}$) instead of one. We assume that each one of these noise sources is in the electromagnetic vacuum state $|0\rangle$. Consequently, the power spectral density of the output noise
\begin{multline}
    \frac{S_\text{noise}(\omega)}{S_{vv}(\omega)} \propto
    \left| T_{n_q}(i\omega) \right|^2 +
    \left| T_{n_q}(-i\omega) \right|^2 \\+
    \left| T_{n_{a1}}(i\omega) \right|^2 +
    \left| T_{n_{a1}}(-i\omega) \right|^2 \\+
    \left| T_{n_{a2}}(i\omega) \right|^2 +
    \left| T_{n_{a2}}(-i\omega) \right|^2.
    \label{eq:PSD_noise}
\end{multline}
This gives the denominator of the right-hand side of Equation~\ref{eq:chisq}.

\section{Optimizing the phase-insensitive gain}
\subsection{Optimal phase-sensitive gain}
\label{app:finding_Gopt}
Our goal in this section is to find the magnitude and phase of $G$ at each frequency that would maximize the signal-to-noise ratio enhancement~\ref{eq:chisq} by the system. We use the second-order approximation of the delay functions (\ref{eq:DMA}), which is the lowest order approximation that accounts for coupled-cavity effects, such as splitting of the resonances induced by optical coupling.

Assuming that $t\sIM,t\sCM\ll1$, we obtain the following analytical solutions for optimal gain magnitude and phase:
\begin{align}
    G_0^{(opt)}(s) &= \left| G\sopt (s) \right| = 1,
    \label{eq:GoptMag}\\
    \varphi_G^{(opt)}(s) &= \arg G\sopt (s) = \frac12 \arctan\left[\frac{s+\gamma_s}{s-\gamma_s}\right],
    \label{eq:GoptPhase}
\end{align}
where $\gamma_s$ is the bandwidth of the sensing cavity (\ref{eq:gamma_s}). The optimal gain magnitude of 1 shows that contribution of the added noise vanishes; the active element actually plays the role of a pure all-pass filter. The role of the introduced phase $\varphi_G$ is to compensate for the unwanted phase shift introduced by the sensing cavity. Combining~(\ref{eq:GoptMag}-\ref{eq:GoptPhase}), we write down the optimal gain $G\sopt$ in the form of (\ref{eq:Gopt}).
\subsection{Sub-optimal gain magnitude}
\label{app:sub_optimal_gain}
\begin{figure}[htb]
    \includegraphics{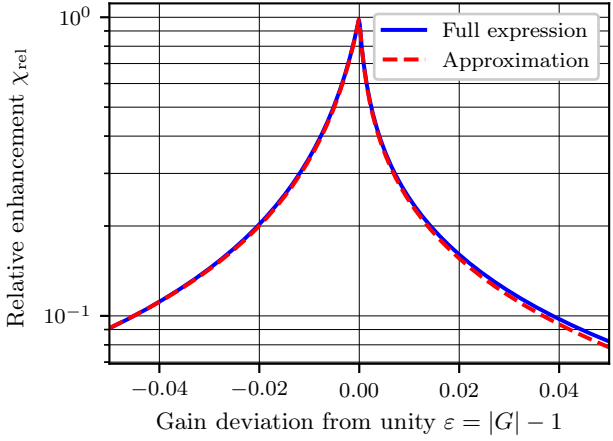}
    \caption{
        Relative SNR enhancement for a sub-optimal filter with gain $G=|G|e^{i\phi_G^{(opt)}}$ if the phase is optimal (given by Eq. \ref{eq:GoptPhase}), but $|G| =1+\varepsilon \ne 1$. Solid line represents the full solution; dashed line shows approximation (\ref{eq:chi_dev_from_opt}).
        \label{fig:enh_vs_gain_magnitude}
    }
\end{figure}

To analyze the PT-symmetric configuration, we need estimations for the sensitivity enhancement if $|G|=G_0$ is not precisely equal to unity, but $\arg G = \phi_G$ is kept equal to the optimal phase $\phi_G^{(opt)}$ as given by (\ref{eq:GoptPhase}). Assuming 
\begin{equation}
    G_0(s) = 1 + \varepsilon(s),
\end{equation}
we write down the relative sensitivity enhancement as
\begin{equation}
    \chi_{\text{rel}}(\omega, \varepsilon) =
    \frac{\chi\left[\omega, G=(1+\varepsilon) e^{i\phi_G^{(opt)}}\right]}
    {\chi\left[\omega, G= e^{i\phi_G^{(opt)}}\right]}.
\end{equation}
Assuming that $|\varepsilon|$ is small, the second-order approximation for the cavity delays (\ref{eq:DMA}) yields
\begin{equation}
    \chi_{\text{rel}}(\omega,\varepsilon)\approx \begin{cases}
        \frac{t\sIM^2}{\sqrt{ t\sIM^4 + 16\varepsilon^2 - 4t\sIM^2 \varepsilon(\varepsilon+2) }}, &\varepsilon < 0, \\
        \frac{t\sIM^2}{\sqrt{ t\sIM^4 + 16\varepsilon^2 + 12t\sIM^2 \varepsilon(\varepsilon+2) }}, &\varepsilon > 0.
    \end{cases}
    \label{eq:chi_dev_from_opt}
\end{equation}
The relative enhancement suppression when moving away from $G_0=1$ does not explicitly depend on frequency. The dependence of this suppression on gain magnitude is shown in Fig.~\ref{fig:enh_vs_gain_magnitude}.

\section{Approximate analytic expressions for sensitivity enhancement}
\label{app:chi_approx}
\begin{figure*}[hbt]
    \centering
    \includegraphics{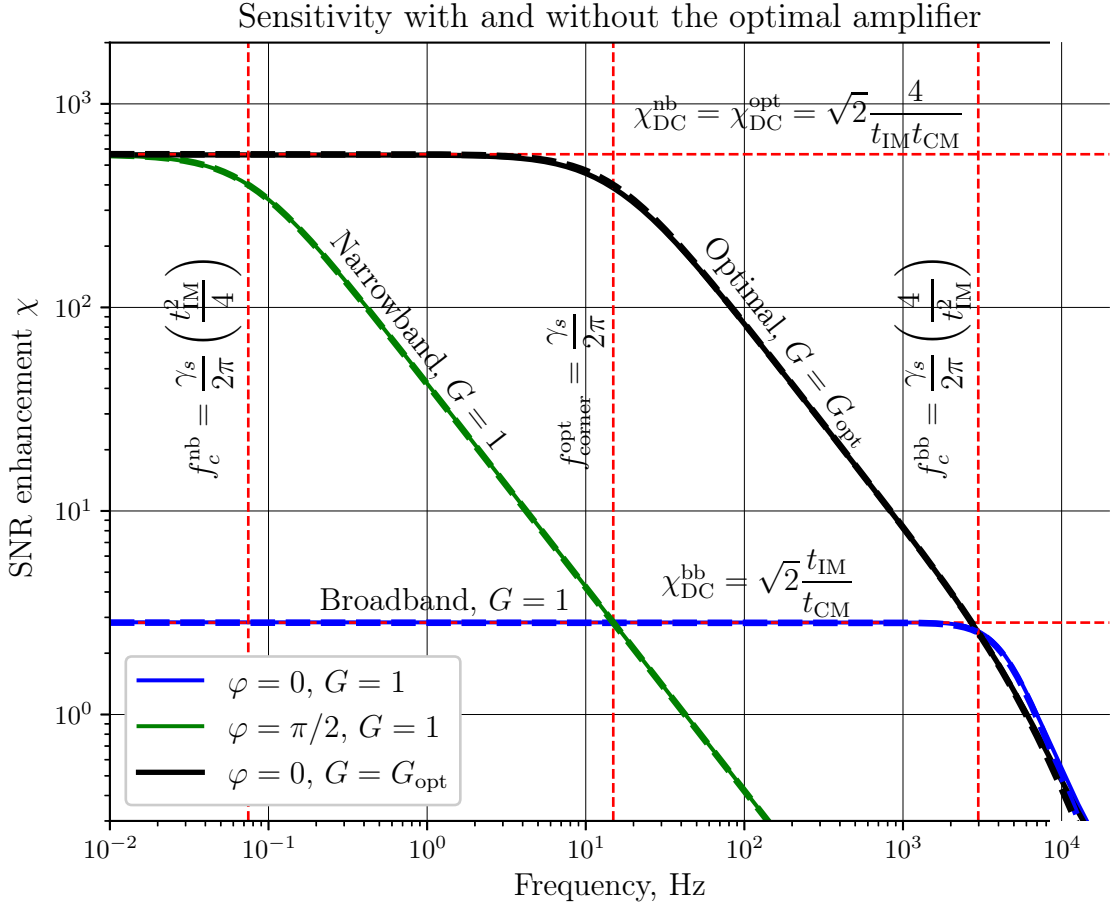}
    \caption{Sensitivity enhancement curves for different configurations: passive narrowband (tuned signal recycling), green; passive broadband (resonant sideband extraction), blue; optimal phase-insensitive filter, black. Solid lines represent full solutions, dashed lines show approximations (\ref{eq:chi_nb}), (\ref{eq:chi_bb_nextlevel}), and (\ref{eq:chi_opt_nextlevel}). DC levels and corner frequencies are shown as thin dashed lines.
    \label{fig:suppl_fig2}}
\end{figure*}

To analyze the key features of phase-insensitive amplification, we obtain simplified analytic expressions for curves that correspond to three distinctive cases. These include two ``passive'' configurations without any phase-insensitive amplification $G\equiv 1$ and two different values of the filter cavity detuning parameter $\varphi$. One of these values, $\varphi=0$, corresponds to the largest bandwidth of the sensitivity curve and the lowest peak sensitivity (we will refer to this as to ``passive broadband'' configuration below); the other, $\varphi=\pi/2$, realizes the passive narrowband configuration characterized by the smallest bandwidth and the highest peak sensitivity. The third configuration assumes optimal filtering ($G=G\sopt$ and $\varphi=0$). To do this, we substitute the second-order expansion of the cavity phase-shifts (\ref{eq:DMA}) to (\ref{eq:PSD_signal},\ref{eq:PSD_noise}). We also expand (\ref{eq:PSD_signal},\ref{eq:PSD_noise}) into second-order series in $t\sIM^2$ and $t\sCM^2$ and neglect the insignificant terms. Results are presented in Fig.~\ref{fig:suppl_fig2} and below. 

\paragraph{Passive, narrowband ($\varphi=\pi/2$, $G\equiv1$).}
Sensitivity enhancement can be approximated as
\begin{equation}
    \chi_{\text{nb}}(\omega)\approx \frac{\XC_{\text{nb}}}{\sqrt{1+\omega^2/\omega_{\text{nb}}^2}},
    \label{eq:chi_nb}
\end{equation}
where $\XC_{\text{nb}}$ is the DC level
\begin{equation}
    \XC_{\text{nb}} = \frac{4\sqrt{2}}{t\sIM t\sCM}
\end{equation}
and $\omega_{\text{nb}}$ is the corner angular frequency
\begin{equation}
    \omega_{\text{nb}} = \gamma_s \frac{t\sIM^2}{4}.
\end{equation}
The integral enhancement is
\begin{equation}
    I_{\text{nb}} = I_0 =
    \int_{0}^{\infty} \chi_\text{nb}^2(\omega)\diff\omega = 
    \frac\pi 2 \XC_{\text{nb}}^2 \omega_{\text{nb}} = \pi/\tau_s
\end{equation}
\paragraph{Passive, broadband ($\varphi=0$, $G\equiv1$)}. Depending on the bandwidth of the filter cavity
\begin{equation}
    \gamma_f = c t\sIM^2/(4L_f),
\end{equation}
there are two different cases. If $\gamma_f\gg \omega_{\text{bb}}$, where $\omega_{\text{bb}}$ is the corner frequency for the broadband case
\begin{equation}
    \omega_{\text{bb}} = \gamma_s \frac{4}{t\sIM^2},
\end{equation}
then the coupled cavity effects can be neglected, and the sensitivity curve is approximated as
\begin{equation}
    \chi_{\text{bb}}(\omega)\approx \frac{\XC_{\text{bb}}}{\sqrt{1+\omega^2/\omega_{\text{bb}}^2}}.
    \label{eq:chi_bb}
\end{equation}
Here the DC level is
\begin{equation}
    \XC_{\text{bb}} = \frac{\sqrt{2}t\sIM}{t\sCM}.
\end{equation}
Since $\XC_{\text{bb}}^2/\XC_{\text{nb}}^2 = \omega_{\text{nb}}/\omega_{\text{bb}}$, for the integral sensitivity we have
\begin{equation}
    I_{\text{bb}}=I_{\text{nb}}=I_0 = \pi/\tau_s.
    \label{eq:Izero}
\end{equation}
We note that one can recover the sensitivity enhancement curve of a single sensing cavity by formally setting $t\sIM=1$ in any of (\ref{eq:chi_nb}) or (\ref{eq:chi_bb}).

If the bandwidth of the filter cavity is comparable to or smaller than the broadband corner frequency, $\gamma_f \lesssim \omega_{\text{bb}}$, then the shape of the sensitivity curve (\ref{eq:chi_bb}) can be significantly distorted by coupled-cavity effects, and a higher-order expansion is required for an accurate representation of the sensitivity in the vicinity of $\omega_{\text{bb}}$:
\begin{equation}
    \chi_{\text{bb}}(\omega)\approx
    \frac{\XC_{\text{bb}}}{ \sqrt{ 
        1 + \frac{\omega^2}{\omega_{\text{bb}}^2} \left[
            1 + \frac{\omega^2 - 2\omega_{\text{bb}}\gamma_f}{\gamma_f^2}
        \right]
    } }.
    \label{eq:chi_bb_nextlevel}
\end{equation}

\paragraph{Optimal filtering ($\varphi=0$, $G=G\sopt$)}. 

In the optimal regime $G=G\sopt$, the DC level of the sensitivity enhancement curve is equal to that of the passive narrowband configuration, but the corner frequency is shifted to $\gamma_s$, i.e. towards higher frequencies:
\begin{equation}
    \chi\sopt(\omega)\approx \frac{\XC_{\text{nb}} }{\sqrt{1+\omega^2/\gamma_s^2}}.
    \label{eq:chi_opt}
\end{equation}
Therefore, the integral sensitivity enhancement is also increased as compared to the passive configurations:
\begin{equation}
    I\sopt  = \frac{4\pi}{\tau_s t\sIM^2} = \frac{4}{t\sIM^2}I_0.
    \label{eq:Iopt}
\end{equation}
Similar to (\ref{eq:chi_bb_nextlevel}), a more accurate approximation of the sensitivity enhancement near $\omega_{\text{bb}}$ is given by
\begin{equation}
    \chi\sopt(\omega)\approx \frac{\XC_{\text{nb}} }{\sqrt{1+\frac{\omega^2}{\gamma_s^2} \left[ 
        1 + \frac{\omega^2}{\gamma_f^2}
    \right]}}.
    \label{eq:chi_opt_nextlevel}
\end{equation}

\section{Nyquist stability analysis}
\label{app:nyquist}
The open-loop transfer function of the system is given by
\begin{equation}
    T_{OL}(s) = r\sIM G^2(s)  Z_f^2(s) r_s(s),
    \label{eq:TOL}
\end{equation}
where
\begin{equation}
    r_{s}(s)=\frac{Z_s^2(s)-r\sCM}{1-r\sCM Z_s^2(s)}
\end{equation}
is the frequency-dependent effective amplitude reflectance off the sensing cavity. 

\begin{figure}[hbt]
    \includegraphics{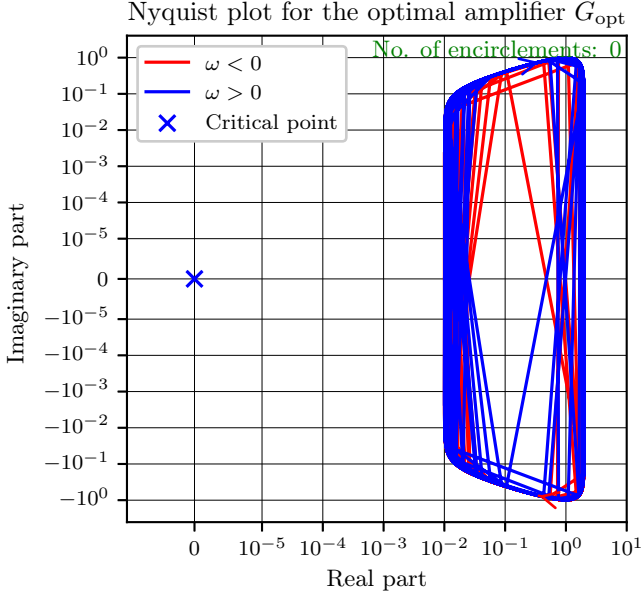}
    \caption{
        Nyquist plot for the optimal phase-insensitive filter $G=G\sopt$. The critical point is encircled zero times; the open-loop gain~(\ref{eq:TOL}) has one unstable pole at $s=\gamma_s$. Therefore, the closed-loop system has a single unstable pole.
        \label{fig:nyq}
    }
\end{figure}

For the optimal filter, $G=G\sopt$ (\ref{eq:Gopt}) and the open-loop function has one unstable pole ($P=1$) at $s=\gamma_s$. The Nyquist plot for this system is shown in Fig.~\ref{fig:nyq}. Since the contour encircles the critical point $N=0$ times, the closed-loop system is also unstable with $Z=N+P=1$ unstable poles. 

The key idea of self-stabilized systems is to replace $G\sopt$ with a transfer function that would make both the open-loop and closed-loop systems stable and, at the same time, would deviate from $G\sopt$ as little as possible, to keep the sensitivity enhancement provided by phase-insensitive amplification. 

\section{PT-symmetric optomechanical filters}
\subsection{Optimal optomechanical coupling}
\label{app:opt_OM_coupling}
We consider a PT-symmetric optomechanical filter: a mechanical oscillator with frequency $\omega_m$ and Q-factor $Q_m$, which is coupled to the circulating field in the filter cavity via the radiation pressure. Pumping the filter cavity with an additional electromagnetic field at the frequency $\omega_0+\omega_m$ with circulating power $P_f$ results in phase-insensitive amplification of the signal~\cite{Li2021,Li2021a} with the gain given by (\ref{eq:GOM}):
\begin{equation*}
    G(s)=G\sOM(s)=1-\frac{
        4g^2\tau_f\omega_m
    }{
        (s+\gamma_m)(s+\gamma_m-2i\omega_m)
    },
\end{equation*}
where $\gamma_m = \omega_m/(2Q_m)$ is the linewidth of the mechanical oscillator and $g\propto\sqrt{P_f}$ is the optomechanical coupling rate~\cite{Aspelmeyer2014}
\begin{equation*}
    g = \sqrt{\frac{16\pi P_f}{m\lambda\omega_m L_f}}.
\end{equation*}
We notice that the open-loop gain (\ref{eq:TOL}) is stable for such a filter. In this section, we show that the value of $g$ suggested in~\cite{Li2021,Li2021a},
\begin{equation}
    g = \frac{\omega_s}{\sqrt{2}} = \frac{1}{\sqrt{2}}\frac{t\sCM}{2\tau_f \tau_s},
\end{equation}
corresponds to the approximation of $G\sopt$ by $G\sPT$ in the operating range of frequencies.

Let us start with considering infinite mechanical Q-factor, i.e. $\gamma_m=0$. Equation (\ref{eq:GOM}) cannot satisfy both conditions (\ref{eq:GoptMag}) and (\ref{eq:GoptPhase}) simultaneously. However, the second condition can be fulfilled in a wide range of frequencies between DC and 
$2\omega_m$, which includes all frequencies between DC and the sensing cavity's FSR. Indeed, assuming $\omega \ll 2\omega_m$, one can write for the gain phase
\begin{equation}
    \varphi_{G}^{(OM)}\approx \arctan\left(-\frac{2 g^2 \tau_f}{\omega}\right).
\end{equation}
Therefore,
\begin{equation}
     \tan 2\varphi_{G}^{(OM)} = \frac{2\tan \varphi_{G}^{(OM)}}{1-\tan^2\varphi_{G}^{(OM)}}
     \approx \frac{4g^2\tau_f\omega}{4g^4 \tau_f^2- \omega^2}.
\end{equation}
On the other hand, it follows from (\ref{eq:GoptPhase}) that
\begin{equation}
    \tan 2\varphi_{G}^{(opt)}=\frac{8 t\sCM^2 \tau_s \omega}{t\sCM^4-16\tau_s^2\omega^2}.
\end{equation}
Comparing these equations, one can find that $\tan 2\varphi_{G}^{(OM)}\approx\tan 2\varphi_{G}^{(opt)}$ when
\begin{equation}
    g^2 = \frac{t\sCM^2}{8\tau_f \tau_s},
\end{equation}
or
\begin{equation}
        g = \omega_s / \sqrt{2}.
\end{equation}
In our schematic, the optomechanical element is placed inside the filter cavity rather than at the input mirror; therefore, the light interacts with it twice per round-trip, hence the required OM coupling rate is reduced by $\sqrt{2}$ as compared to \cite{Li2021,Li2021a}.

\begin{figure}[hbt]
    \includegraphics{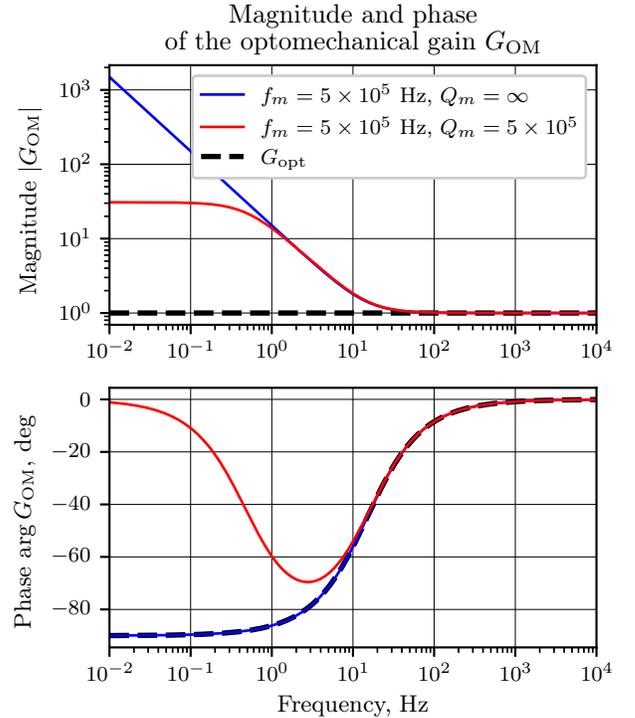}
    \caption{
        Deviation of the optomechanical gain magnitude from unity (top) and optomechanical gain phase (bottom) as functions of frequency in a PT-symmetric optomechanical filter. The two solid curves correspond to a mechanical oscillator without dissipation, $Q_m=\infty$ (blue) and to a mechanical oscillator with $Q_m=5\times 10^5$ (red). The thick dashed line shows the optimal filter with gain~(\ref{eq:Gopt}).
        \label{fig:OM_gain_magphase}
    }
\end{figure}

The phase is shown in Fig.~\ref{fig:OM_gain_magphase}(bottom). One can see that the phase curves for a PT-symmetric filter with infinite $Q_m$ (solid blue) and the optimal filter $G\sopt$ (dashed black) coincide. However, a real mechanical resonator always has a finite Q-factor. The effect of this is illustrated by the black curve, for which $Q_m = 5\times 10^5$. At frequencies below the mechanical linewidth, the phase of $G\sPT$ starts to increase with decreasing frequency and arrives at 0 at DC. This means that the SNR enhancement in such systems is achievable only at frequencies above the mechanical linewidth.

\subsection{Limits of sensitivity enhancement}
\label{app:PT_limits}
Finite mechanical bandwidth is not the only factor limiting the SNR enhancement at low frequencies (LF). The other one comes from (\ref{eq:GoptMag}) and would exist even in a system with infinite mechanical $Q$-factor. Indeed, one can write that the deviation of the gain magnitude from unity, if $Q_m=0$, is
\begin{equation}
    \varepsilon=G_0^{(OM)} -1= \sqrt{1+\eta^2}-1\approx \begin{cases}
        |\eta|, &|\eta| \gg 1,\\
        \eta^2/2, &|\eta| \ll 1,
    \end{cases}
\end{equation}
where (again, assuming that $\omega \ll\omega_m$)
\begin{equation}
    |\eta| \approx \frac{2 g^2 \tau_f}{\omega} = \frac{\omega_s^2 \tau_f}{\omega}=\frac{c t\sCM^2}{4\omega L_s} = \frac{\gamma_s}{\omega}.
\end{equation}
Therefore, for frequencies comparable to or smaller than the bandwidth of the sensing cavity, $\omega\lesssim \gamma_s$, one gets $\varepsilon \gtrsim 1$, and condition (\ref{eq:GoptMag}) is not fulfilled. This is illustrated by Fig.~\ref{fig:OM_gain_magphase} (top).

The LF limit for SNR enhancement imposed by the gain magnitude mismatch can be estimated by substituting $\varepsilon =  |\eta|$ into (\ref{eq:chi_dev_from_opt}). After simplifications, we get equation~\ref{eq:PTSOMchiLF}:
\begin{equation*}
    \chi\sPT(\omega) < \frac{8 t\sIM}{t\sCM^3}\omega\tau_s.
\end{equation*}

\begin{figure}[hbt]%[hbt]
    \centering
    \includegraphics{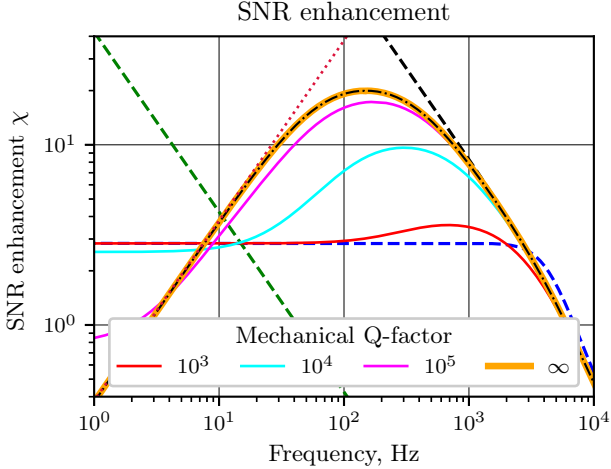}
    \caption{Solid curves represent the sensitivity enhancement by a PT-symmetric optomechanical filter with mechanical resonant frequency $f_m=500$~kHz and different $Q$-factors (as shown in the legend). The thick orange curve corresponds to the limit of sensitivity improvement that can be provided by PT-symmetric filters. The thin black dash-dotted line shows the analytic approximation (\ref{eq:chi_om_infQ}) of the orange curve. The dotted line represents the low-frequency limit (\ref{eq:PTSOMchiLF}) due to the deviation of $|G|$ from unity; black, blue, and green dashed curves are the same as in Fig.~\ref{fig:passive_and_opt}.
    \label{fig:pt_symmetric}}
\end{figure}
This limitation means that there always exists a value of $Q_m$ such that any further decrease of $Q_m$ does not result in any significant sensitivity improvement. This is illustrated in Fig.~\ref{fig:pt_symmetric}, which shows sensitivity enhancement of the optomechanical filter for different quality factors of the mechanical oscillator. The dotted curve represents the limit (\ref{eq:PTSOMchiLF}).

Therefore, sensitivity enhancement by the PT-symmetric optomechanical technique is fundamentally limited at lower frequencies by (\ref{eq:PTSOMchiLF}) and at high frequencies by (\ref{eq:chi_opt}). Using the approximation technique described in Appendix \ref{app:chi_approx}, we approximate $\chi\sPT$ for $Q_m=\infty$ as
\begin{multline}
    \chi\sPT(\omega) \approx \\\frac{8\sqrt{2}t\sCM t\sIM \tau_s\omega}
    { \sqrt{
        t\sCM^8 + 48t\sCM^4 t\sIM^2 \tau_s^2 \omega^2 +
        64\tau_s^4 \omega^4 \left(
            t\sIM^4 +16\tau_f^2 \omega^2
        \right)
    } }.
    \label{eq:chi_om_infQ}
\end{multline}

If the bandwidth of the filter cavity is larger than the frequencies of interest, $\gamma_f \gg \omega_{\text{bb}}$, then the term $16\tau_f\omega^2$ in the denominator of (\ref{eq:chi_om_infQ}) can be neglected, and the integral sensitivity approaches the limit of sensitivity enhancement that can be provided by PT-symmetric optomechanical systems:
\begin{equation}
    I\sPT = \int_{0}^{\infty} \chi\sPT^2(\omega)\diff\omega = 
    \frac{1}{t\sIM} \frac{\pi}{\tau_s} = \frac{1}{t\sIM} I_0 = 
    \frac{t\sIM}{4}I\sopt.
\end{equation}

\begin{table*}[hbt]%The best place to locate the table environment is directly after its first reference in text
\caption{\label{tab:zpk}%
Optimized configurations of gain $G(s) = K \prod_{i=1}^{N_z}(s - z_i) / \prod_{j=1}^{N_p}(s-p_j)$ of the phase-insensitive filter, presented in the form of lists of zeros $z_i$, poles $p_j$, and gain $K$.}
\begin{ruledtabular}
\begin{tabular}{llld{2}@{}l@{}c@{}d{2}@{}ld{2}@{}l@{}c@{}d{2}@{}ld{6}}
\textrm{Configuration}&
\textrm{$\Lambda_f$}&
\textrm{$\Lambda_s$}&
\multicolumn{5}{c}{\textrm{Poles, Hz}}&
\multicolumn{5}{c}{\textrm{Zeros, Hz}}&
\multicolumn{1}{c}{\textrm{Gain}}\\
\colrule
% Optimal controller $G\sopt^2$ (unstable) & 
%  0 & 0 & 
% 14.91 & & & & &
% -14.91 & & & & &
% 1.0\\[3pt]
PT-symmetric filter $G\sPT$
&
0 & 0 & 
-2.5 & $\times 10^{-5}$ & & & &
14.91 & &$+$& i 1.0 & $\times 10^6$ &
 1.0 \\
 ($f_m=500$\,kHz, $Q_m=10^{10}$) &
 &  & 
-2.5 & $\times 10^{-5}$ & $+$ & i 1.0 & $\times 10^6$ &
-14.91 & &$-$& i 2.22 & $\times 10^{-4}$ & \\[3pt]
Optimized with two poles &
0 & 0 &
-1.0 & $\times 10^{-2}$ & $-$ & i3.61 & $\times 10^{-4}$ &
14.82 & &$+$& i17.75 & $\times 10^4$ &
1.019389 \\
(lossless)&
& &
-7.39 & $\times 10^{-1}$ & $+$ & i18.13 & $\times 10^4$ &
-14.79 & &$+$& i5.16 & $\times 10^{-3}$ &
\\[3pt]
Optimized with three poles &
0 & 0 &
-8.56 & & $-$ & i7.48 & $\times 10^{-4}$ &
-21.33 & & & & &
0.998930 \\
(lossless) &
& &
-2.62 & & $+$ & i11.31 & $\times 10^{-5}$ &
-6.91 & & $+$ & i12.78 & &
 \\
 &
& &
-9.33 & & $+$ & i9.90 & $\times 10^{-4}$ &
-6.91 & & $-$ & i12.78 & &
 \\[3pt]
Optimized with two poles &
 $2\times 10^{-3}$ & $5\times 10^{-5}$ &
-1.0 & $\times 10^{-2}$ & $-$ & i2.76 & $\times 10^{-5}$ &
10.84 & & $+$ & i9.98 & $\times 10^5$ &
1.003015 \\
(with loss) &
 & &
-19.36 & $\times 10^{-2}$ & $+$ & i10.03 & $\times 10^5$ &
-14.78 & & $-$ & i3.49 & $\times 10^{-4}$ &
 \\[3pt]
Optimized with three poles &
 $2\times 10^{-3}$ & $5\times 10^{-5}$ &
-7.26 & & $+$ & i8.30 & $\times 10^{-4}$ &
-21.09 & & & & &
0.999188 \\
(with loss) &
 & &
-2.20 & & $+$ & i16.66 & $\times 10^{-5}$ &
-7.18 & & $+$ & i12.30 & &
 \\
 &
 & &
-11.37 & & $+$ & i10.38 & $\times 10^{-4}$ &
-7.18 & & $-$ & i12.30 & &
\end{tabular}
\end{ruledtabular}
\end{table*}

\begin{figure*}[hbtp]
    \includegraphics{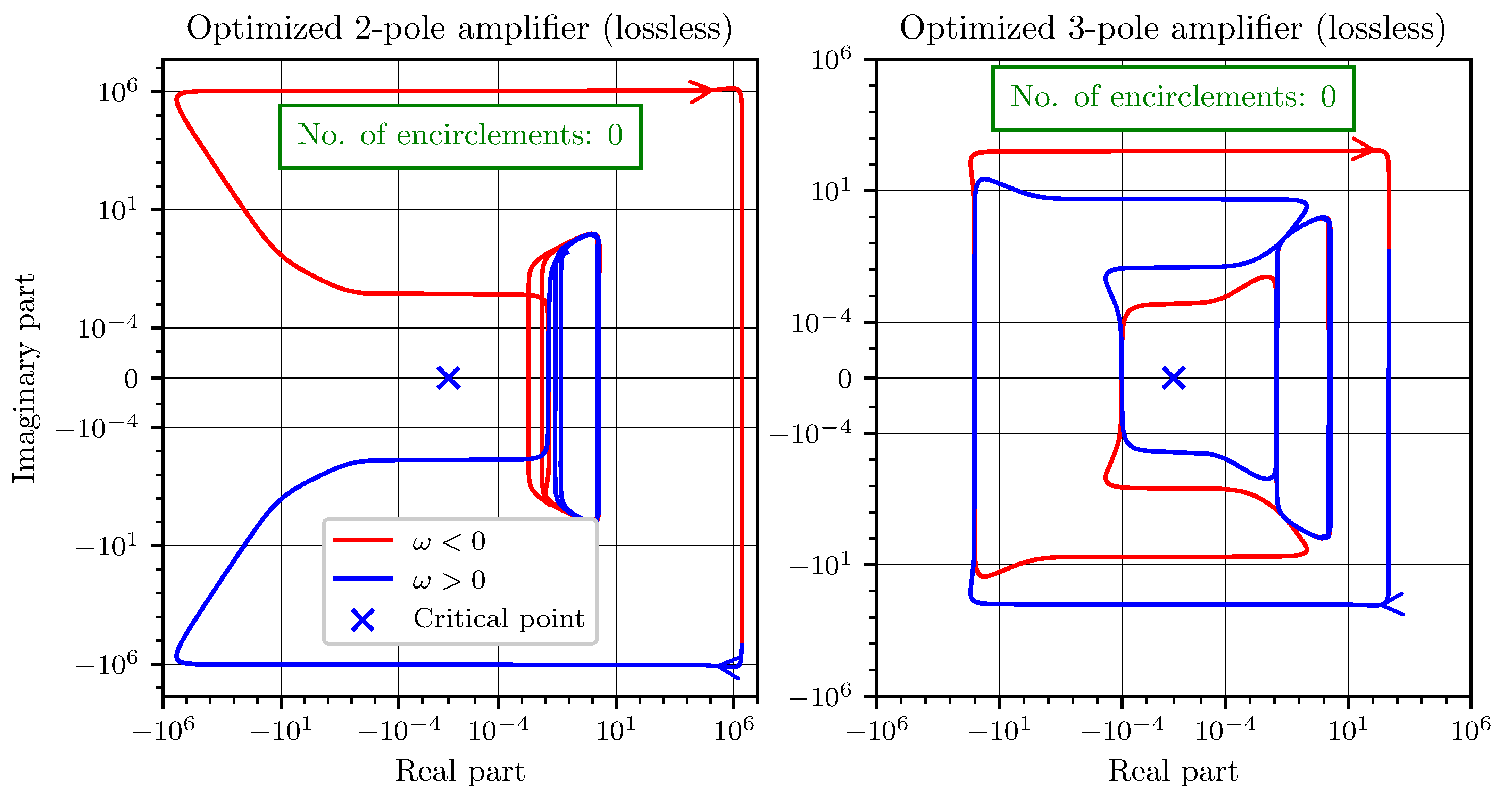}
    \caption{
        Nyquist plots for the systems with optimized two-pole (left) and three-pole (right) phase-insensitive gain $G$. Both systems are stable since the number of times the mapped Nyquist contour encircles the critical point is zero.
        \label{fig:nyqLossless}
    }
\end{figure*}
\section{Optimization details}
\label{app:optimization}
\subsection{Optimization algorithm}
\label{app:optimization_algorithm}
In Section~\ref{sec:constrained_opt}, we present sensitivity curves corresponding to optimized, stable solutions for $G(s)$. In this section, we describe the optimization algorithm we used to find these solutions. 

We optimize $G(s)$ with fixed number of poles $N_p$ and zeros $N_s$ and gain $K$:
\begin{equation}
    G(s) = K \frac{\prod_{i=1}^{N_z}(s - z_i)}
    {\prod_{j=1}^{N_p}(s-p_j)}.
    \label{eq:ZPK}
\end{equation}
We will denote the sets of zeros and poles required to represent a given $G(s)$ as $Z$ and $P$, respectively. The optimization procedure amounts to the minimization of a certain cost function, which depends on $Z$, $P$, and $K$. This minimization is performed via the Nelder-Mead algorithm using standard Python optimization tools.

The cost function is based on the integral sensitivity enhancement by the system in the frequency range between zero and half the FSR of the sensing cavity, $\pi/(2\tau_s)$:
\begin{equation}
    I(Z,P,K,\phi\sLO) = \int_{0}^{\pi/2\tau_s} \chi^2(\omega,Z,P,K,\phi\sLO)\diff\omega.
\end{equation}
The cost function also includes terms whose role is to ensure that both $G$ and the closed-loop system with loop gain (\ref{eq:TOL}) are causal and stable. It is calculated for a given set of zeros $Z$, poles $P$, and gain $K$ as follows:
\begin{enumerate}
    \item By plugging (\ref{eq:ZPK}) into (\ref{eq:PSD_signal}) and (\ref{eq:PSD_noise}) and dividing the former by the latter, we obtain $I$ as a function of given $Z,P,K$ and the homodyne angle $\phi\sLO$.
    \item We maximize $I(Z,P,K,\phi\sLO)$ for the given $Z,P,K$ with respect to $\phi\sLO$ by a separate Nelder-mead optimization procedure, thus obtaining the optimal homodyne angle $\phi\sopt (Z,P,K)$.
    \item We normalize $I(Z,P,K,\phi\sopt)$ by the maximum integral enhancement  $I_0=\pi/\tau_s$ that can be provided by passive systems for which $G\equiv 1$ (see Eq.~\ref{eq:Izero}). The upper limit for the normalized enhancement is given by the quantity~$4/t\sIM^2$ (see Eq.~\ref{eq:Iopt}).
    \item We count the total number $N_{UG}$ of unstable and nearly unstable poles $p_i$ in the set $P$, for which 
    \begin{equation}
        \Re p_i  > -M_{UG},
    \end{equation}
    where $M_{UG}$ is a small positive value describing how close to zero the negative real part of a stable pole can be in our system.
    \item We calculate the total number $N_{UCL}$ of times the mapped Nyquist contour for the system (\ref{eq:TOL}) encircles the critical point $(-1,0)$ clockwise (the winding number). This is done automatically: we introduce a radial coordinate system $(\rho, \varphi)$ with its center at the critical point, unwrap the phase $\varphi$ along the Nyquist contour, and calculate the net amount of full turns by $-360^\circ$ it accumulates along the contour. The system is unstable if $N_{UCL}>0$. We also calculate the minimal distance $\rho\smin$ between the Nyquist contour and the critical point. 
    \item Finally, we calculate the cost function as
    % \begin{multline}
    %     C(Z,P,K) = -\frac{\tau_s}{\pi}I(Z,P,K,\phi\sopt(Z,P,K)) \\
    %     + W\bigg(N_{UG} + N_{UCL} \\ 
    %     +
    %     \left\{\begin{aligned}
    %         0~~&\text{if}~\rho\smin \ge M_\rho \\
    %         1 - \rho\smin/M_\rho~~&\text{if}~\rho\smin < M_\rho
    %     \end{aligned}\right\}
    %     \bigg).
    % \end{multline}
    \begin{widetext}
    \begin{equation}
        C(Z,P,K) = -\frac{\tau_s}{\pi}I(Z,P,K,\phi\sopt(Z,P,K)) + W\left(N_{UG} + N_{UCL} +
        \left\{\begin{aligned}
            0~~&\text{if}~\rho\smin \ge M_\rho \\
            1 - \rho\smin/M_\rho~~&\text{if}~\rho\smin < M_\rho
        \end{aligned}\right\}
        \right).
    \end{equation}
    Here $W$ is the weight of the instability penalty, which is set to a very large number as compared to $4/t\sIM^2$, and $M_\rho$ represents the stability margin -- the measure of how close we allow the Nyquist contour of a stable system to be to the critical point. We used $W=10^8$, $M_{UG}=2\pi\cdot 10^{-2}$\,rad/s, and $M_\rho=10^{-4}$ in our calculations presented in the following sections.
    \end{widetext}
\end{enumerate}
\begin{figure}[bthp]
    \includegraphics{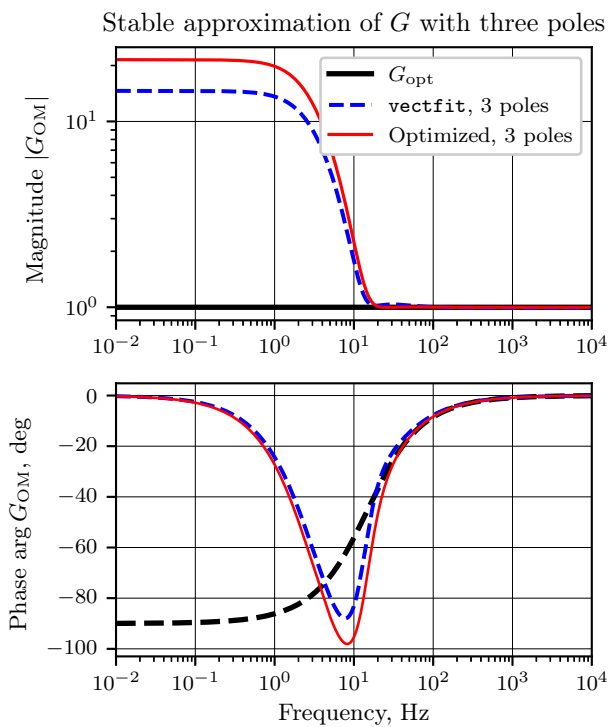}
    \caption{
        The fit of the optimal filter's transfer function $G\sopt$ (thick dashed black curve) with a causal and stable 3-pole transfer function (dashed blue curve), and the optimized 3-pole transfer function (solid red curve).
        \label{fig:fit3poles}
    }
\end{figure}
\subsection{Two-pole solution}
Here we consider a phase-insensitive filter $G$ with $N_p=2$ poles and $N_z=2$ zeros. The poles and zeros are complex, and their real and imaginary parts both act as independent variables in the optimization. Since the optimization space has a large number of dimensions,
\begin{equation}
    N_D = 2N_p + 2N_z + 1,
\end{equation}
and since we do not have a very good measure of how close to stability the closed-loop system is if it is unstable, it is important that the initial guess of $Z,P,K$ resulted in a stable closed-loop system.

For the two-pole system, an obvious choice of the initial guess is the stable PT-symmetric optomechanical solution (\ref{eq:GOM}). The resulting optimized values of $Z,P,K$, which correspond to the cyan sensitivity curve in Fig.~\ref{fig:3pole}, are given in Table~\ref{tab:zpk}. The Nyquist plot of the system is shown in Fig.~\ref{fig:nyqLossless} (left). Note that we shift the critical point to the origin and use the symmetric logarithmic scaling~\footnote{Symmetric logarithmic scaling is logarithmic in both the positive and negative directions from the origin except for a narrow range around zero in which the plot is linear.} for both axes to improve the readability of our Nyquist plots.

\subsection{Three-pole solution}
The integral sensitivity enhancement can be improved by increasing the number of poles and zeros of $G$. In principle, one can approximate $G\sopt$ with a stable $G$ within chosen frequency range with arbitrary precision by choosing a sufficiently large number of poles and zeros. However, this needs to be balanced with the technical complexity and additional losses introduced by the physical realizations of quantum filters, which are likely to increase with the number of poles and zeros. Because of that, in this work, we only show that increasing the number of poles can indeed improve the sensitivity without compromising the stability, by considering the case of $N_p=3$, $N_z=3$.

To get the initial approximation for this case, we fitted $G\sopt$ in the range of angular frequencies $[\gamma_s, 2\pi\cdot 10^5]$\,rad/s with a stable 3-pole $G$. We used our Python implementation\footnote{\url{https://github.com/artemiy-dmitriev/vectfit}} of the \texttt{vectfit} algorithm~\cite{Gustavsen1999} for this. Results of the fit are shown in Fig.~\ref{fig:fit3poles}. Fitted values of $Z,P,K$ were then used as the initial guess for the optimization procedure described in Appendix~\ref{app:optimization_algorithm}. Results of the optimization are given in Table~\ref{tab:zpk} and plotted in Fig.~\ref{fig:fit3poles}, and the corresponding SNR enhancement is shown in Fig.~\ref{fig:3pole} (the magenta curve).
%\nocite{suppl}
\bibliography{qamp}

\begin{thebibliography}{61}%
\makeatletter
\providecommand \@ifxundefined [1]{%
 \@ifx{#1\undefined}
}%
\providecommand \@ifnum [1]{%
 \ifnum #1\expandafter \@firstoftwo
 \else \expandafter \@secondoftwo
 \fi
}%
\providecommand \@ifx [1]{%
 \ifx #1\expandafter \@firstoftwo
 \else \expandafter \@secondoftwo
 \fi
}%
\providecommand \natexlab [1]{#1}%
\providecommand \enquote  [1]{``#1''}%
\providecommand \bibnamefont  [1]{#1}%
\providecommand \bibfnamefont [1]{#1}%
\providecommand \citenamefont [1]{#1}%
\providecommand \href@noop [0]{\@secondoftwo}%
\providecommand \href [0]{\begingroup \@sanitize@url \@href}%
\providecommand \@href[1]{\@@startlink{#1}\@@href}%
\providecommand \@@href[1]{\endgroup#1\@@endlink}%
\providecommand \@sanitize@url [0]{\catcode `\\12\catcode `\$12\catcode
  `\&12\catcode `\#12\catcode `\^12\catcode `\_12\catcode `\%12\relax}%
\providecommand \@@startlink[1]{}%
\providecommand \@@endlink[0]{}%
\providecommand \url  [0]{\begingroup\@sanitize@url \@url }%
\providecommand \@url [1]{\endgroup\@href {#1}{\urlprefix }}%
\providecommand \urlprefix  [0]{URL }%
\providecommand \Eprint [0]{\href }%
\providecommand \doibase [0]{https://doi.org/}%
\providecommand \selectlanguage [0]{\@gobble}%
\providecommand \bibinfo  [0]{\@secondoftwo}%
\providecommand \bibfield  [0]{\@secondoftwo}%
\providecommand \translation [1]{[#1]}%
\providecommand \BibitemOpen [0]{}%
\providecommand \bibitemStop [0]{}%
\providecommand \bibitemNoStop [0]{.\EOS\space}%
\providecommand \EOS [0]{\spacefactor3000\relax}%
\providecommand \BibitemShut  [1]{\csname bibitem#1\endcsname}%
\let\auto@bib@innerbib\@empty
%</preamble>
\bibitem [{\citenamefont {Clerk}\ \emph {et~al.}(2010)\citenamefont {Clerk},
  \citenamefont {Devoret}, \citenamefont {Girvin}, \citenamefont {Marquardt},\
  and\ \citenamefont {Schoelkopf}}]{Clerk2010}%
  \BibitemOpen
  \bibfield  {author} {\bibinfo {author} {\bibfnamefont {A.~A.}\ \bibnamefont
  {Clerk}}, \bibinfo {author} {\bibfnamefont {M.~H.}\ \bibnamefont {Devoret}},
  \bibinfo {author} {\bibfnamefont {S.~M.}\ \bibnamefont {Girvin}}, \bibinfo
  {author} {\bibfnamefont {F.}~\bibnamefont {Marquardt}},\ and\ \bibinfo
  {author} {\bibfnamefont {R.~J.}\ \bibnamefont {Schoelkopf}},\ }\bibfield
  {title} {\bibinfo {title} {Introduction to quantum noise, measurement, and
  amplification},\ }\href {https://doi.org/10.1103/revmodphys.82.1155}
  {\bibfield  {journal} {\bibinfo  {journal} {Reviews of Modern Physics}\
  }\textbf {\bibinfo {volume} {82}},\ \bibinfo {pages} {1155} (\bibinfo {year}
  {2010})}\BibitemShut {NoStop}%
\bibitem [{\citenamefont {Chaudhuri}\ \emph {et~al.}(2018)\citenamefont
  {Chaudhuri}, \citenamefont {Irwin}, \citenamefont {Graham},\ and\
  \citenamefont {Mardon}}]{Chaudhuri2018}%
  \BibitemOpen
  \bibfield  {author} {\bibinfo {author} {\bibfnamefont {S.}~\bibnamefont
  {Chaudhuri}}, \bibinfo {author} {\bibfnamefont {K.}~\bibnamefont {Irwin}},
  \bibinfo {author} {\bibfnamefont {P.~W.}\ \bibnamefont {Graham}},\ and\
  \bibinfo {author} {\bibfnamefont {J.}~\bibnamefont {Mardon}},\ }\bibfield
  {title} {\bibinfo {title} {{Fundamental Limits of Electromagnetic Axion and
  Hidden-Photon Dark Matter Searches: Part I - The Quantum Limit}},\ }\Eprint
  {https://arxiv.org/abs/1803.01627} {arXiv:1803.01627 [hep-ph]}  (\bibinfo
  {year} {2018})\BibitemShut {NoStop}%
\bibitem [{\citenamefont {DeRocco}\ and\ \citenamefont
  {Hook}(2018)}]{DeRocco2018}%
  \BibitemOpen
  \bibfield  {author} {\bibinfo {author} {\bibfnamefont {W.}~\bibnamefont
  {DeRocco}}\ and\ \bibinfo {author} {\bibfnamefont {A.}~\bibnamefont {Hook}},\
  }\bibfield  {title} {\bibinfo {title} {Axion interferometry},\ }\href
  {https://doi.org/10.1103/PhysRevD.98.035021} {\bibfield  {journal} {\bibinfo
  {journal} {Phys. Rev. D}\ }\textbf {\bibinfo {volume} {98}},\ \bibinfo
  {pages} {035021} (\bibinfo {year} {2018})}\BibitemShut {NoStop}%
\bibitem [{\citenamefont {Obata}\ \emph {et~al.}(2018)\citenamefont {Obata},
  \citenamefont {Fujita},\ and\ \citenamefont {Michimura}}]{Obata2018}%
  \BibitemOpen
  \bibfield  {author} {\bibinfo {author} {\bibfnamefont {I.}~\bibnamefont
  {Obata}}, \bibinfo {author} {\bibfnamefont {T.}~\bibnamefont {Fujita}},\ and\
  \bibinfo {author} {\bibfnamefont {Y.}~\bibnamefont {Michimura}},\ }\bibfield
  {title} {\bibinfo {title} {Optical ring cavity search for axion dark
  matter},\ }\href {https://doi.org/10.1103/PhysRevLett.121.161301} {\bibfield
  {journal} {\bibinfo  {journal} {Phys. Rev. Lett.}\ }\textbf {\bibinfo
  {volume} {121}},\ \bibinfo {pages} {161301} (\bibinfo {year}
  {2018})}\BibitemShut {NoStop}%
\bibitem [{\citenamefont {Liu}\ \emph {et~al.}(2019)\citenamefont {Liu},
  \citenamefont {Elwood}, \citenamefont {Evans},\ and\ \citenamefont
  {Thaler}}]{Liu2019}%
  \BibitemOpen
  \bibfield  {author} {\bibinfo {author} {\bibfnamefont {H.}~\bibnamefont
  {Liu}}, \bibinfo {author} {\bibfnamefont {B.~D.}\ \bibnamefont {Elwood}},
  \bibinfo {author} {\bibfnamefont {M.}~\bibnamefont {Evans}},\ and\ \bibinfo
  {author} {\bibfnamefont {J.}~\bibnamefont {Thaler}},\ }\bibfield  {title}
  {\bibinfo {title} {Searching for axion dark matter with birefringent
  cavities},\ }\href {https://doi.org/10.1103/PhysRevD.100.023548} {\bibfield
  {journal} {\bibinfo  {journal} {Phys. Rev. D}\ }\textbf {\bibinfo {volume}
  {100}},\ \bibinfo {pages} {023548} (\bibinfo {year} {2019})}\BibitemShut
  {NoStop}%
\bibitem [{\citenamefont {Martynov}\ and\ \citenamefont
  {Miao}(2020)}]{Martynov2020}%
  \BibitemOpen
  \bibfield  {author} {\bibinfo {author} {\bibfnamefont {D.}~\bibnamefont
  {Martynov}}\ and\ \bibinfo {author} {\bibfnamefont {H.}~\bibnamefont
  {Miao}},\ }\bibfield  {title} {\bibinfo {title} {Quantum-enhanced
  interferometry for axion searches},\ }\href
  {https://doi.org/10.1103/PhysRevD.101.095034} {\bibfield  {journal} {\bibinfo
   {journal} {Phys. Rev. D}\ }\textbf {\bibinfo {volume} {101}},\ \bibinfo
  {pages} {095034} (\bibinfo {year} {2020})}\BibitemShut {NoStop}%
\bibitem [{\citenamefont {Aspelmeyer}\ \emph {et~al.}(2014)\citenamefont
  {Aspelmeyer}, \citenamefont {Kippenberg},\ and\ \citenamefont
  {Marquardt}}]{Aspelmeyer2014}%
  \BibitemOpen
  \bibfield  {author} {\bibinfo {author} {\bibfnamefont {M.}~\bibnamefont
  {Aspelmeyer}}, \bibinfo {author} {\bibfnamefont {T.~J.}\ \bibnamefont
  {Kippenberg}},\ and\ \bibinfo {author} {\bibfnamefont {F.}~\bibnamefont
  {Marquardt}},\ }\bibfield  {title} {\bibinfo {title} {Cavity optomechanics},\
  }\href {https://doi.org/10.1103/revmodphys.86.1391} {\bibfield  {journal}
  {\bibinfo  {journal} {Reviews of Modern Physics}\ }\textbf {\bibinfo {volume}
  {86}},\ \bibinfo {pages} {1391} (\bibinfo {year} {2014})}\BibitemShut
  {NoStop}%
\bibitem [{\citenamefont {Aasi}\ \emph {et~al.}(2015)\citenamefont {Aasi},
  \citenamefont {Abbott}, \citenamefont {Abbott}, \citenamefont {Abbott},
  \citenamefont {Abernathy}, \citenamefont {Ackley} \emph
  {et~al.}}]{Fritschel2015}%
  \BibitemOpen
  \bibfield  {author} {\bibinfo {author} {\bibfnamefont {J.}~\bibnamefont
  {Aasi}}, \bibinfo {author} {\bibfnamefont {B.~P.}\ \bibnamefont {Abbott}},
  \bibinfo {author} {\bibfnamefont {R.}~\bibnamefont {Abbott}}, \bibinfo
  {author} {\bibfnamefont {T.}~\bibnamefont {Abbott}}, \bibinfo {author}
  {\bibfnamefont {M.~R.}\ \bibnamefont {Abernathy}}, \bibinfo {author}
  {\bibfnamefont {K.}~\bibnamefont {Ackley}}, \emph {et~al.},\ }\bibfield
  {title} {\bibinfo {title} {Advanced {LIGO}},\ }\href
  {https://doi.org/10.1088/0264-9381/32/7/074001} {\bibfield  {journal}
  {\bibinfo  {journal} {Classical and Quantum Gravity}\ }\textbf {\bibinfo
  {volume} {32}},\ \bibinfo {pages} {074001} (\bibinfo {year}
  {2015})}\BibitemShut {NoStop}%
\bibitem [{\citenamefont {Acernese}\ \emph {et~al.}(2014)\citenamefont
  {Acernese}, \citenamefont {Agathos}, \citenamefont {Agatsuma}, \citenamefont
  {Aisa}, \citenamefont {Allemandou}, \citenamefont {Allocca} \emph
  {et~al.}}]{Acernese2014}%
  \BibitemOpen
  \bibfield  {author} {\bibinfo {author} {\bibfnamefont {F.}~\bibnamefont
  {Acernese}}, \bibinfo {author} {\bibfnamefont {M.}~\bibnamefont {Agathos}},
  \bibinfo {author} {\bibfnamefont {K.}~\bibnamefont {Agatsuma}}, \bibinfo
  {author} {\bibfnamefont {D.}~\bibnamefont {Aisa}}, \bibinfo {author}
  {\bibfnamefont {N.}~\bibnamefont {Allemandou}}, \bibinfo {author}
  {\bibfnamefont {A.}~\bibnamefont {Allocca}}, \emph {et~al.},\ }\bibfield
  {title} {\bibinfo {title} {{Advanced Virgo: a second-generation
  interferometric gravitational wave detector}},\ }\href
  {https://doi.org/10.1088/0264-9381/32/2/024001} {\bibfield  {journal}
  {\bibinfo  {journal} {Classical and Quantum Gravity}\ }\textbf {\bibinfo
  {volume} {32}},\ \bibinfo {pages} {024001} (\bibinfo {year}
  {2014})}\BibitemShut {NoStop}%
\bibitem [{\citenamefont {Caves}\ and\ \citenamefont
  {Schumaker}(1985)}]{Caves1985}%
  \BibitemOpen
  \bibfield  {author} {\bibinfo {author} {\bibfnamefont {C.~M.}\ \bibnamefont
  {Caves}}\ and\ \bibinfo {author} {\bibfnamefont {B.~L.}\ \bibnamefont
  {Schumaker}},\ }\bibfield  {title} {\bibinfo {title} {{New formalism for
  two-photon quantum optics. I. Quadrature phases and squeezed states}},\
  }\href {https://doi.org/10.1103/physreva.31.3068} {\bibfield  {journal}
  {\bibinfo  {journal} {Physical Review A}\ }\textbf {\bibinfo {volume} {31}},\
  \bibinfo {pages} {3068} (\bibinfo {year} {1985})}\BibitemShut {NoStop}%
\bibitem [{\citenamefont {Schumaker}\ and\ \citenamefont
  {Caves}(1985)}]{Schumaker1985}%
  \BibitemOpen
  \bibfield  {author} {\bibinfo {author} {\bibfnamefont {B.~L.}\ \bibnamefont
  {Schumaker}}\ and\ \bibinfo {author} {\bibfnamefont {C.~M.}\ \bibnamefont
  {Caves}},\ }\bibfield  {title} {\bibinfo {title} {{New formalism for
  two-photon quantum optics. {II}. Mathematical foundation and compact
  notation}},\ }\href {https://doi.org/10.1103/physreva.31.3093} {\bibfield
  {journal} {\bibinfo  {journal} {Physical Review A}\ }\textbf {\bibinfo
  {volume} {31}},\ \bibinfo {pages} {3093} (\bibinfo {year}
  {1985})}\BibitemShut {NoStop}%
\bibitem [{\citenamefont {Kimble}\ \emph {et~al.}(2001)\citenamefont {Kimble},
  \citenamefont {Levin}, \citenamefont {Matsko}, \citenamefont {Thorne},\ and\
  \citenamefont {Vyatchanin}}]{Kimble2001}%
  \BibitemOpen
  \bibfield  {author} {\bibinfo {author} {\bibfnamefont {H.~J.}\ \bibnamefont
  {Kimble}}, \bibinfo {author} {\bibfnamefont {Y.}~\bibnamefont {Levin}},
  \bibinfo {author} {\bibfnamefont {A.~B.}\ \bibnamefont {Matsko}}, \bibinfo
  {author} {\bibfnamefont {K.~S.}\ \bibnamefont {Thorne}},\ and\ \bibinfo
  {author} {\bibfnamefont {S.~P.}\ \bibnamefont {Vyatchanin}},\ }\bibfield
  {title} {\bibinfo {title} {Conversion of conventional gravitational-wave
  interferometers into quantum nondemolition interferometers by modifying their
  input and/or output optics},\ }\href
  {https://doi.org/10.1103/physrevd.65.022002} {\bibfield  {journal} {\bibinfo
  {journal} {Physical Review D}\ }\textbf {\bibinfo {volume} {65}},\ \bibinfo
  {pages} {022002} (\bibinfo {year} {2001})}\BibitemShut {NoStop}%
\bibitem [{\citenamefont {Martynov}\ \emph {et~al.}(2016)\citenamefont
  {Martynov}, \citenamefont {Hall}, \citenamefont {Abbott}, \citenamefont
  {Abbott}, \citenamefont {Abbott}, \citenamefont {Adams} \emph
  {et~al.}}]{Martynov2016}%
  \BibitemOpen
  \bibfield  {author} {\bibinfo {author} {\bibfnamefont {D.~V.}\ \bibnamefont
  {Martynov}}, \bibinfo {author} {\bibfnamefont {E.~D.}\ \bibnamefont {Hall}},
  \bibinfo {author} {\bibfnamefont {B.~P.}\ \bibnamefont {Abbott}}, \bibinfo
  {author} {\bibfnamefont {R.}~\bibnamefont {Abbott}}, \bibinfo {author}
  {\bibfnamefont {T.~D.}\ \bibnamefont {Abbott}}, \bibinfo {author}
  {\bibfnamefont {C.}~\bibnamefont {Adams}}, \emph {et~al.},\ }\bibfield
  {title} {\bibinfo {title} {{Sensitivity of the Advanced {LIGO} detectors at
  the beginning of gravitational wave astronomy}},\ }\href
  {https://doi.org/10.1103/physrevd.93.112004} {\bibfield  {journal} {\bibinfo
  {journal} {Physical Review D}\ }\textbf {\bibinfo {volume} {93}},\ \bibinfo
  {pages} {112004} (\bibinfo {year} {2016})}\BibitemShut {NoStop}%
\bibitem [{\citenamefont {Buikema}\ \emph {et~al.}(2020)\citenamefont
  {Buikema}, \citenamefont {Cahillane}, \citenamefont {Mansell}, \citenamefont
  {Blair}, \citenamefont {Abbott}, \citenamefont {Adams} \emph
  {et~al.}}]{Buikema2020}%
  \BibitemOpen
  \bibfield  {author} {\bibinfo {author} {\bibfnamefont {A.}~\bibnamefont
  {Buikema}}, \bibinfo {author} {\bibfnamefont {C.}~\bibnamefont {Cahillane}},
  \bibinfo {author} {\bibfnamefont {G.~L.}\ \bibnamefont {Mansell}}, \bibinfo
  {author} {\bibfnamefont {C.~D.}\ \bibnamefont {Blair}}, \bibinfo {author}
  {\bibfnamefont {R.}~\bibnamefont {Abbott}}, \bibinfo {author} {\bibfnamefont
  {C.}~\bibnamefont {Adams}}, \emph {et~al.},\ }\bibfield  {title} {\bibinfo
  {title} {Sensitivity and performance of the {A}dvanced {LIGO} detectors in
  the third observing run},\ }\href
  {https://doi.org/10.1103/physrevd.102.062003} {\bibfield  {journal} {\bibinfo
   {journal} {Physical Review D}\ }\textbf {\bibinfo {volume} {102}},\ \bibinfo
  {pages} {062003} (\bibinfo {year} {2020})}\BibitemShut {NoStop}%
\bibitem [{\citenamefont {Reitze}\ \emph {et~al.}(2019)\citenamefont {Reitze}
  \emph {et~al.}}]{Reitze2019}%
  \BibitemOpen
  \bibfield  {author} {\bibinfo {author} {\bibfnamefont {D.}~\bibnamefont
  {Reitze}} \emph {et~al.},\ }\bibfield  {title} {\bibinfo {title} {{Cosmic
  Explorer: The U.S. Contribution to Gravitational-Wave Astronomy beyond
  LIGO}},\ }\Eprint {https://arxiv.org/abs/1907.04833} {1907.04833}  (\bibinfo
  {year} {2019})\BibitemShut {NoStop}%
\bibitem [{\citenamefont {{ET Editorial Team}}(2020)}]{ET2020}%
  \BibitemOpen
  \bibfield  {author} {\bibinfo {author} {\bibnamefont {{ET Editorial Team}}},\
  }\href {https://apps.et-gw.eu/tds/ql/?c=15418} {\emph {\bibinfo {title}
  {{E}instein {T}elescope {D}esign {R}eport {U}pdate}}},\ \bibinfo {type}
  {Tech. Rep.}\ (\bibinfo {year} {2020})\BibitemShut {NoStop}%
\bibitem [{\citenamefont {Klimenko}\ \emph {et~al.}(2011)\citenamefont
  {Klimenko}, \citenamefont {Vedovato}, \citenamefont {Drago}, \citenamefont
  {Mazzolo}, \citenamefont {Mitselmakher}, \citenamefont {Pankow},
  \citenamefont {Prodi}, \citenamefont {Re}, \citenamefont {Salemi},\ and\
  \citenamefont {Yakushin}}]{Klimenko2011}%
  \BibitemOpen
  \bibfield  {author} {\bibinfo {author} {\bibfnamefont {S.}~\bibnamefont
  {Klimenko}}, \bibinfo {author} {\bibfnamefont {G.}~\bibnamefont {Vedovato}},
  \bibinfo {author} {\bibfnamefont {M.}~\bibnamefont {Drago}}, \bibinfo
  {author} {\bibfnamefont {G.}~\bibnamefont {Mazzolo}}, \bibinfo {author}
  {\bibfnamefont {G.}~\bibnamefont {Mitselmakher}}, \bibinfo {author}
  {\bibfnamefont {C.}~\bibnamefont {Pankow}}, \bibinfo {author} {\bibfnamefont
  {G.}~\bibnamefont {Prodi}}, \bibinfo {author} {\bibfnamefont
  {V.}~\bibnamefont {Re}}, \bibinfo {author} {\bibfnamefont {F.}~\bibnamefont
  {Salemi}},\ and\ \bibinfo {author} {\bibfnamefont {I.}~\bibnamefont
  {Yakushin}},\ }\bibfield  {title} {\bibinfo {title} {Localization of
  gravitational wave sources with networks of advanced detectors},\ }\href
  {https://doi.org/10.1103/PhysRevD.83.102001} {\bibfield  {journal} {\bibinfo
  {journal} {Phys. Rev. D}\ }\textbf {\bibinfo {volume} {83}},\ \bibinfo
  {pages} {102001} (\bibinfo {year} {2011})}\BibitemShut {NoStop}%
\bibitem [{\citenamefont {Martynov}\ \emph {et~al.}(2019)\citenamefont
  {Martynov}, \citenamefont {Miao}, \citenamefont {Yang}, \citenamefont
  {Vivanco}, \citenamefont {Thrane}, \citenamefont {Smith} \emph
  {et~al.}}]{Martynov2019}%
  \BibitemOpen
  \bibfield  {author} {\bibinfo {author} {\bibfnamefont {D.}~\bibnamefont
  {Martynov}}, \bibinfo {author} {\bibfnamefont {H.}~\bibnamefont {Miao}},
  \bibinfo {author} {\bibfnamefont {H.}~\bibnamefont {Yang}}, \bibinfo {author}
  {\bibfnamefont {F.~H.}\ \bibnamefont {Vivanco}}, \bibinfo {author}
  {\bibfnamefont {E.}~\bibnamefont {Thrane}}, \bibinfo {author} {\bibfnamefont
  {R.}~\bibnamefont {Smith}}, \emph {et~al.},\ }\bibfield  {title} {\bibinfo
  {title} {Exploring the sensitivity of gravitational wave detectors to neutron
  star physics},\ }\href {https://doi.org/10.1103/physrevd.99.102004}
  {\bibfield  {journal} {\bibinfo  {journal} {Physical Review D}\ }\textbf
  {\bibinfo {volume} {99}},\ \bibinfo {pages} {102004} (\bibinfo {year}
  {2019})}\BibitemShut {NoStop}%
\bibitem [{\citenamefont {Miao}\ \emph {et~al.}(2018)\citenamefont {Miao},
  \citenamefont {Yang},\ and\ \citenamefont {Martynov}}]{Miao2018}%
  \BibitemOpen
  \bibfield  {author} {\bibinfo {author} {\bibfnamefont {H.}~\bibnamefont
  {Miao}}, \bibinfo {author} {\bibfnamefont {H.}~\bibnamefont {Yang}},\ and\
  \bibinfo {author} {\bibfnamefont {D.}~\bibnamefont {Martynov}},\ }\bibfield
  {title} {\bibinfo {title} {Towards the design of gravitational-wave detectors
  for probing neutron-star physics},\ }\href
  {https://doi.org/10.1103/physrevd.98.044044} {\bibfield  {journal} {\bibinfo
  {journal} {Physical Review D}\ }\textbf {\bibinfo {volume} {98}},\ \bibinfo
  {pages} {044044} (\bibinfo {year} {2018})}\BibitemShut {NoStop}%
\bibitem [{\citenamefont {Berti}\ \emph {et~al.}(2016)\citenamefont {Berti},
  \citenamefont {Sesana}, \citenamefont {Barausse}, \citenamefont {Cardoso},\
  and\ \citenamefont {Belczynski}}]{Berti2016}%
  \BibitemOpen
  \bibfield  {author} {\bibinfo {author} {\bibfnamefont {E.}~\bibnamefont
  {Berti}}, \bibinfo {author} {\bibfnamefont {A.}~\bibnamefont {Sesana}},
  \bibinfo {author} {\bibfnamefont {E.}~\bibnamefont {Barausse}}, \bibinfo
  {author} {\bibfnamefont {V.}~\bibnamefont {Cardoso}},\ and\ \bibinfo {author}
  {\bibfnamefont {K.}~\bibnamefont {Belczynski}},\ }\bibfield  {title}
  {\bibinfo {title} {Spectroscopy of kerr black holes with earth- and
  space-based interferometers},\ }\href
  {https://doi.org/10.1103/PhysRevLett.117.101102} {\bibfield  {journal}
  {\bibinfo  {journal} {Phys. Rev. Lett.}\ }\textbf {\bibinfo {volume} {117}},\
  \bibinfo {pages} {101102} (\bibinfo {year} {2016})}\BibitemShut {NoStop}%
\bibitem [{\citenamefont {Laghi}\ \emph {et~al.}(2021)\citenamefont {Laghi},
  \citenamefont {Carullo}, \citenamefont {Veitch},\ and\ \citenamefont
  {Pozzo}}]{Laghi2021}%
  \BibitemOpen
  \bibfield  {author} {\bibinfo {author} {\bibfnamefont {D.}~\bibnamefont
  {Laghi}}, \bibinfo {author} {\bibfnamefont {G.}~\bibnamefont {Carullo}},
  \bibinfo {author} {\bibfnamefont {J.}~\bibnamefont {Veitch}},\ and\ \bibinfo
  {author} {\bibfnamefont {W.~D.}\ \bibnamefont {Pozzo}},\ }\bibfield  {title}
  {\bibinfo {title} {Quantum black hole spectroscopy: probing the quantum
  nature of the black hole area using {LIGO}{\textendash}virgo ringdown
  detections},\ }\href {https://doi.org/10.1088/1361-6382/abde19} {\bibfield
  {journal} {\bibinfo  {journal} {Classical and Quantum Gravity}\ }\textbf
  {\bibinfo {volume} {38}},\ \bibinfo {pages} {095005} (\bibinfo {year}
  {2021})}\BibitemShut {NoStop}%
\bibitem [{\citenamefont {Meers}(1988)}]{Meers1988}%
  \BibitemOpen
  \bibfield  {author} {\bibinfo {author} {\bibfnamefont {B.~J.}\ \bibnamefont
  {Meers}},\ }\bibfield  {title} {\bibinfo {title} {Recycling in
  laser-interferometric gravitational-wave detectors},\ }\href
  {https://doi.org/10.1103/physrevd.38.2317} {\bibfield  {journal} {\bibinfo
  {journal} {Physical Review D}\ }\textbf {\bibinfo {volume} {38}},\ \bibinfo
  {pages} {2317} (\bibinfo {year} {1988})}\BibitemShut {NoStop}%
\bibitem [{\citenamefont {Mizuno}\ \emph {et~al.}(1993)\citenamefont {Mizuno},
  \citenamefont {Strain}, \citenamefont {Nelson}, \citenamefont {Chen},
  \citenamefont {Schilling}, \citenamefont {Rüdiger}, \citenamefont
  {Winkler},\ and\ \citenamefont {Danzmann}}]{Mizuno1993}%
  \BibitemOpen
  \bibfield  {author} {\bibinfo {author} {\bibfnamefont {J.}~\bibnamefont
  {Mizuno}}, \bibinfo {author} {\bibfnamefont {K.}~\bibnamefont {Strain}},
  \bibinfo {author} {\bibfnamefont {P.}~\bibnamefont {Nelson}}, \bibinfo
  {author} {\bibfnamefont {J.}~\bibnamefont {Chen}}, \bibinfo {author}
  {\bibfnamefont {R.}~\bibnamefont {Schilling}}, \bibinfo {author}
  {\bibfnamefont {A.}~\bibnamefont {Rüdiger}}, \bibinfo {author}
  {\bibfnamefont {W.}~\bibnamefont {Winkler}},\ and\ \bibinfo {author}
  {\bibfnamefont {K.}~\bibnamefont {Danzmann}},\ }\bibfield  {title} {\bibinfo
  {title} {Resonant sideband extraction: a new configuration for
  interferometric gravitational wave detectors},\ }\href
  {https://doi.org/10.1016/0375-9601(93)90620-f} {\bibfield  {journal}
  {\bibinfo  {journal} {Physics Letters A}\ }\textbf {\bibinfo {volume}
  {175}},\ \bibinfo {pages} {273} (\bibinfo {year} {1993})}\BibitemShut
  {NoStop}%
\bibitem [{\citenamefont {Caves}(1981)}]{Caves1981}%
  \BibitemOpen
  \bibfield  {author} {\bibinfo {author} {\bibfnamefont {C.~M.}\ \bibnamefont
  {Caves}},\ }\bibfield  {title} {\bibinfo {title} {Quantum-mechanical noise in
  an interferometer},\ }\href {https://doi.org/10.1103/physrevd.23.1693}
  {\bibfield  {journal} {\bibinfo  {journal} {Physical Review D}\ }\textbf
  {\bibinfo {volume} {23}},\ \bibinfo {pages} {1693} (\bibinfo {year}
  {1981})}\BibitemShut {NoStop}%
\bibitem [{\citenamefont {Schnabel}(2017)}]{Schnabel2017}%
  \BibitemOpen
  \bibfield  {author} {\bibinfo {author} {\bibfnamefont {R.}~\bibnamefont
  {Schnabel}},\ }\bibfield  {title} {\bibinfo {title} {Squeezed states of light
  and their applications in laser interferometers},\ }\href
  {https://doi.org/10.1016/j.physrep.2017.04.001} {\bibfield  {journal}
  {\bibinfo  {journal} {Physics Reports}\ }\textbf {\bibinfo {volume} {684}},\
  \bibinfo {pages} {1} (\bibinfo {year} {2017})}\BibitemShut {NoStop}%
\bibitem [{\citenamefont {Buonanno}\ and\ \citenamefont
  {Chen}(2002)}]{Buonanno2002}%
  \BibitemOpen
  \bibfield  {author} {\bibinfo {author} {\bibfnamefont {A.}~\bibnamefont
  {Buonanno}}\ and\ \bibinfo {author} {\bibfnamefont {Y.}~\bibnamefont
  {Chen}},\ }\bibfield  {title} {\bibinfo {title} {Signal recycled
  laser-interferometer gravitational-wave detectors as optical springs},\
  }\href {https://doi.org/10.1103/physrevd.65.042001} {\bibfield  {journal}
  {\bibinfo  {journal} {Physical Review D}\ }\textbf {\bibinfo {volume} {65}},\
  \bibinfo {pages} {042001} (\bibinfo {year} {2002})}\BibitemShut {NoStop}%
\bibitem [{\citenamefont {Korobko}\ \emph {et~al.}(2019)\citenamefont
  {Korobko}, \citenamefont {Ma}, \citenamefont {Chen},\ and\ \citenamefont
  {Schnabel}}]{Korobko2019}%
  \BibitemOpen
  \bibfield  {author} {\bibinfo {author} {\bibfnamefont {M.}~\bibnamefont
  {Korobko}}, \bibinfo {author} {\bibfnamefont {Y.}~\bibnamefont {Ma}},
  \bibinfo {author} {\bibfnamefont {Y.}~\bibnamefont {Chen}},\ and\ \bibinfo
  {author} {\bibfnamefont {R.}~\bibnamefont {Schnabel}},\ }\bibfield  {title}
  {\bibinfo {title} {Quantum expander for gravitational-wave observatories},\
  }\href {https://doi.org/10.1038/s41377-019-0230-2} {\bibfield  {journal}
  {\bibinfo  {journal} {Light: Science {\&} Applications}\ }\textbf {\bibinfo
  {volume} {8}},\ \bibinfo {pages} {118} (\bibinfo {year} {2019})}\BibitemShut
  {NoStop}%
\bibitem [{\citenamefont {Brooks}\ \emph {et~al.}(2021)\citenamefont {Brooks},
  \citenamefont {Vajente}, \citenamefont {Yamamoto}, \citenamefont {Abbott},
  \citenamefont {Adams}, \citenamefont {Adhikari} \emph {et~al.}}]{Brooks2021}%
  \BibitemOpen
  \bibfield  {author} {\bibinfo {author} {\bibfnamefont {A.~F.}\ \bibnamefont
  {Brooks}}, \bibinfo {author} {\bibfnamefont {G.}~\bibnamefont {Vajente}},
  \bibinfo {author} {\bibfnamefont {H.}~\bibnamefont {Yamamoto}}, \bibinfo
  {author} {\bibfnamefont {R.}~\bibnamefont {Abbott}}, \bibinfo {author}
  {\bibfnamefont {C.}~\bibnamefont {Adams}}, \bibinfo {author} {\bibfnamefont
  {R.~X.}\ \bibnamefont {Adhikari}}, \emph {et~al.},\ }\bibfield  {title}
  {\bibinfo {title} {{Point absorbers in Advanced {LIGO}}},\ }\href
  {https://doi.org/10.1364/ao.419689} {\bibfield  {journal} {\bibinfo
  {journal} {Applied Optics}\ }\textbf {\bibinfo {volume} {60}},\ \bibinfo
  {pages} {4047} (\bibinfo {year} {2021})}\BibitemShut {NoStop}%
\bibitem [{\citenamefont {Braginsky}\ \emph {et~al.}(2001)\citenamefont
  {Braginsky}, \citenamefont {Strigin},\ and\ \citenamefont
  {Vyatchanin}}]{Braginsky2001}%
  \BibitemOpen
  \bibfield  {author} {\bibinfo {author} {\bibfnamefont {V.}~\bibnamefont
  {Braginsky}}, \bibinfo {author} {\bibfnamefont {S.}~\bibnamefont {Strigin}},\
  and\ \bibinfo {author} {\bibfnamefont {S.}~\bibnamefont {Vyatchanin}},\
  }\bibfield  {title} {\bibinfo {title} {{Parametric oscillatory instability in
  Fabry{\textendash}Perot interferometer}},\ }\href
  {https://doi.org/10.1016/s0375-9601(01)00510-2} {\bibfield  {journal}
  {\bibinfo  {journal} {Physics Letters A}\ }\textbf {\bibinfo {volume}
  {287}},\ \bibinfo {pages} {331} (\bibinfo {year} {2001})}\BibitemShut
  {NoStop}%
\bibitem [{\citenamefont {Chen}\ \emph {et~al.}(2015)\citenamefont {Chen},
  \citenamefont {Zhao}, \citenamefont {Danilishin}, \citenamefont {Ju},
  \citenamefont {Blair}, \citenamefont {Wang} \emph {et~al.}}]{Chen2015}%
  \BibitemOpen
  \bibfield  {author} {\bibinfo {author} {\bibfnamefont {X.}~\bibnamefont
  {Chen}}, \bibinfo {author} {\bibfnamefont {C.}~\bibnamefont {Zhao}}, \bibinfo
  {author} {\bibfnamefont {S.}~\bibnamefont {Danilishin}}, \bibinfo {author}
  {\bibfnamefont {L.}~\bibnamefont {Ju}}, \bibinfo {author} {\bibfnamefont
  {D.}~\bibnamefont {Blair}}, \bibinfo {author} {\bibfnamefont
  {H.}~\bibnamefont {Wang}}, \emph {et~al.},\ }\bibfield  {title} {\bibinfo
  {title} {Observation of three-mode parametric instability},\ }\href
  {https://doi.org/10.1103/physreva.91.033832} {\bibfield  {journal} {\bibinfo
  {journal} {Physical Review A}\ }\textbf {\bibinfo {volume} {91}},\ \bibinfo
  {pages} {033832} (\bibinfo {year} {2015})}\BibitemShut {NoStop}%
\bibitem [{\citenamefont {Evans}\ \emph {et~al.}(2015)\citenamefont {Evans},
  \citenamefont {Gras}, \citenamefont {Fritschel}, \citenamefont {Miller},
  \citenamefont {Barsotti}, \citenamefont {Martynov} \emph
  {et~al.}}]{Evans2015}%
  \BibitemOpen
  \bibfield  {author} {\bibinfo {author} {\bibfnamefont {M.}~\bibnamefont
  {Evans}}, \bibinfo {author} {\bibfnamefont {S.}~\bibnamefont {Gras}},
  \bibinfo {author} {\bibfnamefont {P.}~\bibnamefont {Fritschel}}, \bibinfo
  {author} {\bibfnamefont {J.}~\bibnamefont {Miller}}, \bibinfo {author}
  {\bibfnamefont {L.}~\bibnamefont {Barsotti}}, \bibinfo {author}
  {\bibfnamefont {D.}~\bibnamefont {Martynov}}, \emph {et~al.},\ }\bibfield
  {title} {\bibinfo {title} {{Observation of Parametric Instability in Advanced
  {LIGO}}},\ }\href {https://doi.org/10.1103/physrevlett.114.161102} {\bibfield
   {journal} {\bibinfo  {journal} {Physical Review Letters}\ }\textbf {\bibinfo
  {volume} {114}},\ \bibinfo {pages} {161102} (\bibinfo {year}
  {2015})}\BibitemShut {NoStop}%
\bibitem [{\citenamefont {Braginsky}\ \emph {et~al.}(1992)\citenamefont
  {Braginsky}, \citenamefont {Khalili},\ and\ \citenamefont
  {Thorne}}]{Braginsky1992}%
  \BibitemOpen
  \bibfield  {author} {\bibinfo {author} {\bibfnamefont {V.~B.}\ \bibnamefont
  {Braginsky}}, \bibinfo {author} {\bibfnamefont {F.~Y.}\ \bibnamefont
  {Khalili}},\ and\ \bibinfo {author} {\bibfnamefont {K.~S.}\ \bibnamefont
  {Thorne}},\ }\href {https://doi.org/10.1017/cbo9780511622748} {\emph
  {\bibinfo {title} {Quantum Measurement}}}\ (\bibinfo  {publisher} {Cambridge
  University Press},\ \bibinfo {year} {1992})\BibitemShut {NoStop}%
\bibitem [{\citenamefont {Braginsky}(2000)}]{Braginsky2000}%
  \BibitemOpen
  \bibfield  {author} {\bibinfo {author} {\bibfnamefont {V.~B.}\ \bibnamefont
  {Braginsky}},\ }\bibfield  {title} {\bibinfo {title} {Energetic quantum limit
  in large-scale interferometers},\ }in\ \href
  {https://doi.org/10.1063/1.1291855} {\emph {\bibinfo {booktitle} {{AIP}
  Conference Proceedings}}}\ (\bibinfo  {publisher} {{AIP}},\ \bibinfo {year}
  {2000})\BibitemShut {NoStop}%
\bibitem [{\citenamefont {Tsang}\ \emph {et~al.}(2011)\citenamefont {Tsang},
  \citenamefont {Wiseman},\ and\ \citenamefont {Caves}}]{Tsang2011}%
  \BibitemOpen
  \bibfield  {author} {\bibinfo {author} {\bibfnamefont {M.}~\bibnamefont
  {Tsang}}, \bibinfo {author} {\bibfnamefont {H.~M.}\ \bibnamefont {Wiseman}},\
  and\ \bibinfo {author} {\bibfnamefont {C.~M.}\ \bibnamefont {Caves}},\
  }\bibfield  {title} {\bibinfo {title} {Fundamental quantum limit to waveform
  estimation},\ }\href {https://doi.org/10.1103/physrevlett.106.090401}
  {\bibfield  {journal} {\bibinfo  {journal} {Physical Review Letters}\
  }\textbf {\bibinfo {volume} {106}},\ \bibinfo {pages} {090401} (\bibinfo
  {year} {2011})}\BibitemShut {NoStop}%
\bibitem [{\citenamefont {Grote}\ \emph {et~al.}(2013)\citenamefont {Grote},
  \citenamefont {Danzmann}, \citenamefont {Dooley}, \citenamefont {Schnabel},
  \citenamefont {Slutsky},\ and\ \citenamefont {Vahlbruch}}]{Grote2013}%
  \BibitemOpen
  \bibfield  {author} {\bibinfo {author} {\bibfnamefont {H.}~\bibnamefont
  {Grote}}, \bibinfo {author} {\bibfnamefont {K.}~\bibnamefont {Danzmann}},
  \bibinfo {author} {\bibfnamefont {K.~L.}\ \bibnamefont {Dooley}}, \bibinfo
  {author} {\bibfnamefont {R.}~\bibnamefont {Schnabel}}, \bibinfo {author}
  {\bibfnamefont {J.}~\bibnamefont {Slutsky}},\ and\ \bibinfo {author}
  {\bibfnamefont {H.}~\bibnamefont {Vahlbruch}},\ }\bibfield  {title} {\bibinfo
  {title} {First long-term application of squeezed states of light in a
  gravitational-wave observatory},\ }\href
  {https://doi.org/10.1103/physrevlett.110.181101} {\bibfield  {journal}
  {\bibinfo  {journal} {Physical Review Letters}\ }\textbf {\bibinfo {volume}
  {110}},\ \bibinfo {pages} {181101} (\bibinfo {year} {2013})}\BibitemShut
  {NoStop}%
\bibitem [{\citenamefont {Aasi}\ \emph {et~al.}(2013)\citenamefont {Aasi},
  \citenamefont {Abadie}, \citenamefont {Abbott}, \citenamefont {Abbott},
  \citenamefont {Abbott}, \citenamefont {Abernathy} \emph {et~al.}}]{Aasi2013}%
  \BibitemOpen
  \bibfield  {author} {\bibinfo {author} {\bibfnamefont {J.}~\bibnamefont
  {Aasi}}, \bibinfo {author} {\bibfnamefont {J.}~\bibnamefont {Abadie}},
  \bibinfo {author} {\bibfnamefont {B.~P.}\ \bibnamefont {Abbott}}, \bibinfo
  {author} {\bibfnamefont {R.}~\bibnamefont {Abbott}}, \bibinfo {author}
  {\bibfnamefont {T.~D.}\ \bibnamefont {Abbott}}, \bibinfo {author}
  {\bibfnamefont {M.~R.}\ \bibnamefont {Abernathy}}, \emph {et~al.},\
  }\bibfield  {title} {\bibinfo {title} {Enhanced sensitivity of the {LIGO}
  gravitational wave detector by using squeezed states of light},\ }\href
  {https://doi.org/10.1038/nphoton.2013.177} {\bibfield  {journal} {\bibinfo
  {journal} {Nature Photonics}\ }\textbf {\bibinfo {volume} {7}},\ \bibinfo
  {pages} {613} (\bibinfo {year} {2013})}\BibitemShut {NoStop}%
\bibitem [{\citenamefont {Acernese}\ \emph {et~al.}(2019)\citenamefont
  {Acernese}, \citenamefont {Agathos}, \citenamefont {Aiello}, \citenamefont
  {Allocca}, \citenamefont {Amato}, \citenamefont {Ansoldi} \emph
  {et~al.}}]{Acernese2019}%
  \BibitemOpen
  \bibfield  {author} {\bibinfo {author} {\bibfnamefont {F.}~\bibnamefont
  {Acernese}}, \bibinfo {author} {\bibfnamefont {M.}~\bibnamefont {Agathos}},
  \bibinfo {author} {\bibfnamefont {L.}~\bibnamefont {Aiello}}, \bibinfo
  {author} {\bibfnamefont {A.}~\bibnamefont {Allocca}}, \bibinfo {author}
  {\bibfnamefont {A.}~\bibnamefont {Amato}}, \bibinfo {author} {\bibfnamefont
  {S.}~\bibnamefont {Ansoldi}}, \emph {et~al.},\ }\bibfield  {title} {\bibinfo
  {title} {{Increasing the Astrophysical Reach of the Advanced Virgo Detector
  via the Application of Squeezed Vacuum States of Light}},\ }\href
  {https://doi.org/10.1103/physrevlett.123.231108} {\bibfield  {journal}
  {\bibinfo  {journal} {Physical Review Letters}\ }\textbf {\bibinfo {volume}
  {123}},\ \bibinfo {pages} {231108} (\bibinfo {year} {2019})}\BibitemShut
  {NoStop}%
\bibitem [{\citenamefont {Miao}\ \emph {et~al.}(2019)\citenamefont {Miao},
  \citenamefont {Smith},\ and\ \citenamefont {Evans}}]{Miao2019}%
  \BibitemOpen
  \bibfield  {author} {\bibinfo {author} {\bibfnamefont {H.}~\bibnamefont
  {Miao}}, \bibinfo {author} {\bibfnamefont {N.~D.}\ \bibnamefont {Smith}},\
  and\ \bibinfo {author} {\bibfnamefont {M.}~\bibnamefont {Evans}},\ }\bibfield
   {title} {\bibinfo {title} {Quantum limit for laser interferometric
  gravitational-wave detectors from optical dissipation},\ }\href
  {https://doi.org/10.1103/physrevx.9.011053} {\bibfield  {journal} {\bibinfo
  {journal} {Physical Review X}\ }\textbf {\bibinfo {volume} {9}},\ \bibinfo
  {pages} {011053} (\bibinfo {year} {2019})}\BibitemShut {NoStop}%
\bibitem [{\citenamefont {Caves}(1982)}]{Caves1982}%
  \BibitemOpen
  \bibfield  {author} {\bibinfo {author} {\bibfnamefont {C.~M.}\ \bibnamefont
  {Caves}},\ }\bibfield  {title} {\bibinfo {title} {Quantum limits on noise in
  linear amplifiers},\ }\href {https://doi.org/10.1103/physrevd.26.1817}
  {\bibfield  {journal} {\bibinfo  {journal} {Physical Review D}\ }\textbf
  {\bibinfo {volume} {26}},\ \bibinfo {pages} {1817} (\bibinfo {year}
  {1982})}\BibitemShut {NoStop}%
\bibitem [{Note101()}]{Note101}%
  \BibitemOpen
  \bibinfo {note} {The word ``amplification'' has been a common term for the
  phase-insensitive processes since the Caves' paper~\cite {Caves1982}.
  However, Caves' formalism (\ref {eq:ph-ins-amp}) is applicable to systems
  with any gain $G$, including those with $|G|\le 1$ (e.g. attenuators and
  filters), which do not ``amplify'' the signal. For most phase-insensitive
  elements considered in this work, the gain magnitude $|G|$ is either equal to
  unity or is very close to it; for this reason, to avoid confusion, we use the
  term ``filters'', rather than ``amplifiers'', when referring to such
  systems.}\BibitemShut {Stop}%
\bibitem [{\citenamefont {Miao}\ \emph {et~al.}(2015)\citenamefont {Miao},
  \citenamefont {Ma}, \citenamefont {Zhao},\ and\ \citenamefont
  {Chen}}]{Miao2015}%
  \BibitemOpen
  \bibfield  {author} {\bibinfo {author} {\bibfnamefont {H.}~\bibnamefont
  {Miao}}, \bibinfo {author} {\bibfnamefont {Y.}~\bibnamefont {Ma}}, \bibinfo
  {author} {\bibfnamefont {C.}~\bibnamefont {Zhao}},\ and\ \bibinfo {author}
  {\bibfnamefont {Y.}~\bibnamefont {Chen}},\ }\bibfield  {title} {\bibinfo
  {title} {Enhancing the bandwidth of gravitational-wave detectors with
  unstable optomechanical filters},\ }\href
  {https://doi.org/10.1103/physrevlett.115.211104} {\bibfield  {journal}
  {\bibinfo  {journal} {Physical Review Letters}\ }\textbf {\bibinfo {volume}
  {115}},\ \bibinfo {pages} {211104} (\bibinfo {year} {2015})}\BibitemShut
  {NoStop}%
\bibitem [{\citenamefont {Page}\ \emph {et~al.}(2018)\citenamefont {Page},
  \citenamefont {Qin}, \citenamefont {LaFontaine}, \citenamefont {Zhao},
  \citenamefont {Ju},\ and\ \citenamefont {Blair}}]{Page2018}%
  \BibitemOpen
  \bibfield  {author} {\bibinfo {author} {\bibfnamefont {M.}~\bibnamefont
  {Page}}, \bibinfo {author} {\bibfnamefont {J.}~\bibnamefont {Qin}}, \bibinfo
  {author} {\bibfnamefont {J.}~\bibnamefont {LaFontaine}}, \bibinfo {author}
  {\bibfnamefont {C.}~\bibnamefont {Zhao}}, \bibinfo {author} {\bibfnamefont
  {L.}~\bibnamefont {Ju}},\ and\ \bibinfo {author} {\bibfnamefont
  {D.}~\bibnamefont {Blair}},\ }\bibfield  {title} {\bibinfo {title} {Enhanced
  detection of high frequency gravitational waves using optically diluted
  optomechanical filters},\ }\href {https://doi.org/10.1103/physrevd.97.124060}
  {\bibfield  {journal} {\bibinfo  {journal} {Physical Review D}\ }\textbf
  {\bibinfo {volume} {97}},\ \bibinfo {pages} {124060} (\bibinfo {year}
  {2018})}\BibitemShut {NoStop}%
\bibitem [{\citenamefont {Bentley}\ \emph {et~al.}(2019)\citenamefont
  {Bentley}, \citenamefont {Jones}, \citenamefont {Martynov}, \citenamefont
  {Freise},\ and\ \citenamefont {Miao}}]{Bentley2019}%
  \BibitemOpen
  \bibfield  {author} {\bibinfo {author} {\bibfnamefont {J.}~\bibnamefont
  {Bentley}}, \bibinfo {author} {\bibfnamefont {P.}~\bibnamefont {Jones}},
  \bibinfo {author} {\bibfnamefont {D.}~\bibnamefont {Martynov}}, \bibinfo
  {author} {\bibfnamefont {A.}~\bibnamefont {Freise}},\ and\ \bibinfo {author}
  {\bibfnamefont {H.}~\bibnamefont {Miao}},\ }\bibfield  {title} {\bibinfo
  {title} {Converting the signal-recycling cavity into an unstable
  optomechanical filter to enhance the detection bandwidth of
  gravitational-wave detectors},\ }\href
  {https://doi.org/10.1103/physrevd.99.102001} {\bibfield  {journal} {\bibinfo
  {journal} {Physical Review D}\ }\textbf {\bibinfo {volume} {99}},\ \bibinfo
  {pages} {102001} (\bibinfo {year} {2019})}\BibitemShut {NoStop}%
\bibitem [{\citenamefont {Page}\ \emph {et~al.}(2021)\citenamefont {Page},
  \citenamefont {Goryachev}, \citenamefont {Miao}, \citenamefont {Chen},
  \citenamefont {Ma}, \citenamefont {Mason} \emph {et~al.}}]{Page2021}%
  \BibitemOpen
  \bibfield  {author} {\bibinfo {author} {\bibfnamefont {M.~A.}\ \bibnamefont
  {Page}}, \bibinfo {author} {\bibfnamefont {M.}~\bibnamefont {Goryachev}},
  \bibinfo {author} {\bibfnamefont {H.}~\bibnamefont {Miao}}, \bibinfo {author}
  {\bibfnamefont {Y.}~\bibnamefont {Chen}}, \bibinfo {author} {\bibfnamefont
  {Y.}~\bibnamefont {Ma}}, \bibinfo {author} {\bibfnamefont {D.}~\bibnamefont
  {Mason}}, \emph {et~al.},\ }\bibfield  {title} {\bibinfo {title}
  {Gravitational wave detectors with broadband high frequency sensitivity},\
  }\href {https://doi.org/10.1038/s42005-021-00526-2} {\bibfield  {journal}
  {\bibinfo  {journal} {Communications Physics}\ }\textbf {\bibinfo {volume}
  {4}},\ \bibinfo {pages} {27} (\bibinfo {year} {2021})}\BibitemShut {NoStop}%
\bibitem [{\citenamefont {Zhang}\ \emph
  {et~al.}(2021{\natexlab{a}})\citenamefont {Zhang}, \citenamefont {Bentley},\
  and\ \citenamefont {Miao}}]{Zhang2021}%
  \BibitemOpen
  \bibfield  {author} {\bibinfo {author} {\bibfnamefont {T.}~\bibnamefont
  {Zhang}}, \bibinfo {author} {\bibfnamefont {J.}~\bibnamefont {Bentley}},\
  and\ \bibinfo {author} {\bibfnamefont {H.}~\bibnamefont {Miao}},\ }\bibfield
  {title} {\bibinfo {title} {A broadband signal recycling scheme for
  approaching the quantum limit from optical losses},\ }\href
  {https://doi.org/10.3390/galaxies9010003} {\bibfield  {journal} {\bibinfo
  {journal} {Galaxies}\ }\textbf {\bibinfo {volume} {9}},\ \bibinfo {pages} {3}
  (\bibinfo {year} {2021}{\natexlab{a}})}\BibitemShut {NoStop}%
\bibitem [{\citenamefont {Zhang}\ \emph
  {et~al.}(2021{\natexlab{b}})\citenamefont {Zhang}, \citenamefont {Smetana},
  \citenamefont {Chen}, \citenamefont {Bentley}, \citenamefont {Martynov},
  \citenamefont {Miao}, \citenamefont {East},\ and\ \citenamefont
  {Yang}}]{Zhang2021a}%
  \BibitemOpen
  \bibfield  {author} {\bibinfo {author} {\bibfnamefont {T.}~\bibnamefont
  {Zhang}}, \bibinfo {author} {\bibfnamefont {J.}~\bibnamefont {Smetana}},
  \bibinfo {author} {\bibfnamefont {Y.}~\bibnamefont {Chen}}, \bibinfo {author}
  {\bibfnamefont {J.}~\bibnamefont {Bentley}}, \bibinfo {author} {\bibfnamefont
  {D.}~\bibnamefont {Martynov}}, \bibinfo {author} {\bibfnamefont
  {H.}~\bibnamefont {Miao}}, \bibinfo {author} {\bibfnamefont {W.~E.}\
  \bibnamefont {East}},\ and\ \bibinfo {author} {\bibfnamefont
  {H.}~\bibnamefont {Yang}},\ }\bibfield  {title} {\bibinfo {title} {Toward
  observing neutron star collapse with gravitational wave detectors},\ }\href
  {https://doi.org/10.1103/physrevd.103.044063} {\bibfield  {journal} {\bibinfo
   {journal} {Physical Review D}\ }\textbf {\bibinfo {volume} {103}},\ \bibinfo
  {pages} {044063} (\bibinfo {year} {2021}{\natexlab{b}})}\BibitemShut
  {NoStop}%
\bibitem [{\citenamefont {Li}\ \emph {et~al.}(2020)\citenamefont {Li},
  \citenamefont {Goryachev}, \citenamefont {Ma}, \citenamefont {Tobar},
  \citenamefont {Zhao}, \citenamefont {Adhikari},\ and\ \citenamefont
  {Chen}}]{Li2021}%
  \BibitemOpen
  \bibfield  {author} {\bibinfo {author} {\bibfnamefont {X.}~\bibnamefont
  {Li}}, \bibinfo {author} {\bibfnamefont {M.}~\bibnamefont {Goryachev}},
  \bibinfo {author} {\bibfnamefont {Y.}~\bibnamefont {Ma}}, \bibinfo {author}
  {\bibfnamefont {M.~E.}\ \bibnamefont {Tobar}}, \bibinfo {author}
  {\bibfnamefont {C.}~\bibnamefont {Zhao}}, \bibinfo {author} {\bibfnamefont
  {R.~X.}\ \bibnamefont {Adhikari}},\ and\ \bibinfo {author} {\bibfnamefont
  {Y.}~\bibnamefont {Chen}},\ }\bibfield  {title} {\bibinfo {title} {{Broadband
  sensitivity improvement via coherent quantum feedback with PT symmetry}},\
  }\Eprint {https://arxiv.org/abs/2012.00836} {arXiv:2012.00836 [quant-ph]}
  (\bibinfo {year} {2020})\BibitemShut {NoStop}%
\bibitem [{\citenamefont {Li}\ \emph {et~al.}(2021)\citenamefont {Li},
  \citenamefont {Smetana}, \citenamefont {Ubhi}, \citenamefont {Bentley},
  \citenamefont {Chen}, \citenamefont {Ma}, \citenamefont {Miao},\ and\
  \citenamefont {Martynov}}]{Li2021a}%
  \BibitemOpen
  \bibfield  {author} {\bibinfo {author} {\bibfnamefont {X.}~\bibnamefont
  {Li}}, \bibinfo {author} {\bibfnamefont {J.}~\bibnamefont {Smetana}},
  \bibinfo {author} {\bibfnamefont {A.~S.}\ \bibnamefont {Ubhi}}, \bibinfo
  {author} {\bibfnamefont {J.}~\bibnamefont {Bentley}}, \bibinfo {author}
  {\bibfnamefont {Y.}~\bibnamefont {Chen}}, \bibinfo {author} {\bibfnamefont
  {Y.}~\bibnamefont {Ma}}, \bibinfo {author} {\bibfnamefont {H.}~\bibnamefont
  {Miao}},\ and\ \bibinfo {author} {\bibfnamefont {D.}~\bibnamefont
  {Martynov}},\ }\bibfield  {title} {\bibinfo {title} {{Enhancing
  interferometer sensitivity without sacrificing bandwidth and stability:
  Beyond single-mode and resolved-sideband approximation}},\ }\href
  {https://doi.org/10.1103/physrevd.103.122001} {\bibfield  {journal} {\bibinfo
   {journal} {Phys. Rev. D}\ }\textbf {\bibinfo {volume} {103}},\ \bibinfo
  {pages} {122001} (\bibinfo {year} {2021})}\BibitemShut {NoStop}%
\bibitem [{\citenamefont {Özdemir}\ \emph {et~al.}(2019)\citenamefont
  {Özdemir}, \citenamefont {Rotter}, \citenamefont {Nori},\ and\ \citenamefont
  {Yang}}]{Oezdemir2019}%
  \BibitemOpen
  \bibfield  {author} {\bibinfo {author} {\bibfnamefont {{\c{S}}.~K.}\
  \bibnamefont {Özdemir}}, \bibinfo {author} {\bibfnamefont {S.}~\bibnamefont
  {Rotter}}, \bibinfo {author} {\bibfnamefont {F.}~\bibnamefont {Nori}},\ and\
  \bibinfo {author} {\bibfnamefont {L.}~\bibnamefont {Yang}},\ }\bibfield
  {title} {\bibinfo {title} {Parity{\textendash}time symmetry and exceptional
  points in photonics},\ }\href {https://doi.org/10.1038/s41563-019-0304-9}
  {\bibfield  {journal} {\bibinfo  {journal} {Nature Materials}\ }\textbf
  {\bibinfo {volume} {18}},\ \bibinfo {pages} {783} (\bibinfo {year}
  {2019})}\BibitemShut {NoStop}%
\bibitem [{\citenamefont {Buonanno}\ and\ \citenamefont
  {Chen}(2001)}]{Buonanno2001}%
  \BibitemOpen
  \bibfield  {author} {\bibinfo {author} {\bibfnamefont {A.}~\bibnamefont
  {Buonanno}}\ and\ \bibinfo {author} {\bibfnamefont {Y.}~\bibnamefont
  {Chen}},\ }\bibfield  {title} {\bibinfo {title} {Quantum noise in second
  generation, signal-recycled laser interferometric gravitational-wave
  detectors},\ }\href {https://doi.org/10.1103/physrevd.64.042006} {\bibfield
  {journal} {\bibinfo  {journal} {Physical Review D}\ }\textbf {\bibinfo
  {volume} {64}},\ \bibinfo {pages} {042006} (\bibinfo {year}
  {2001})}\BibitemShut {NoStop}%
\bibitem [{\citenamefont {{\AA}str{\"o}m}\ and\ \citenamefont
  {Murray}(2021)}]{Aastrom2021}%
  \BibitemOpen
  \bibfield  {author} {\bibinfo {author} {\bibfnamefont {K.~J.}\ \bibnamefont
  {{\AA}str{\"o}m}}\ and\ \bibinfo {author} {\bibfnamefont {R.~M.}\
  \bibnamefont {Murray}},\ }\href@noop {} {\emph {\bibinfo {title} {{Feedback
  Systems: An Introduction for Scientists and Engineers, Second Edition}}}}\
  (\bibinfo  {publisher} {{Princeton University Press}},\ \bibinfo {year}
  {2021})\BibitemShut {NoStop}%
\bibitem [{\citenamefont {Baratchart}\ \emph {et~al.}(1997)\citenamefont
  {Baratchart}, \citenamefont {Leblond}, \citenamefont {Partington},\ and\
  \citenamefont {Torkhani}}]{Baratchart1997}%
  \BibitemOpen
  \bibfield  {author} {\bibinfo {author} {\bibfnamefont {L.}~\bibnamefont
  {Baratchart}}, \bibinfo {author} {\bibfnamefont {J.}~\bibnamefont {Leblond}},
  \bibinfo {author} {\bibfnamefont {J.}~\bibnamefont {Partington}},\ and\
  \bibinfo {author} {\bibfnamefont {N.}~\bibnamefont {Torkhani}},\ }\bibfield
  {title} {\bibinfo {title} {Robust identification from band-limited data},\
  }\href {https://doi.org/10.1109/9.623101} {\bibfield  {journal} {\bibinfo
  {journal} {{IEEE} Transactions on Automatic Control}\ }\textbf {\bibinfo
  {volume} {42}},\ \bibinfo {pages} {1318} (\bibinfo {year}
  {1997})}\BibitemShut {NoStop}%
\bibitem [{\citenamefont {Gustavsen}\ and\ \citenamefont
  {Semlyen}(1999)}]{Gustavsen1999}%
  \BibitemOpen
  \bibfield  {author} {\bibinfo {author} {\bibfnamefont {B.}~\bibnamefont
  {Gustavsen}}\ and\ \bibinfo {author} {\bibfnamefont {A.}~\bibnamefont
  {Semlyen}},\ }\bibfield  {title} {\bibinfo {title} {Rational approximation of
  frequency domain responses by vector fitting},\ }\href
  {https://doi.org/10.1109/61.772353} {\bibfield  {journal} {\bibinfo
  {journal} {{IEEE} Transactions on Power Delivery}\ }\textbf {\bibinfo
  {volume} {14}},\ \bibinfo {pages} {1052} (\bibinfo {year}
  {1999})}\BibitemShut {NoStop}%
\bibitem [{\citenamefont {Dwyer}\ \emph {et~al.}(2013)\citenamefont {Dwyer},
  \citenamefont {Barsotti}, \citenamefont {Chua}, \citenamefont {Evans},
  \citenamefont {Factourovich}, \citenamefont {Gustafson} \emph
  {et~al.}}]{Dwyer2013}%
  \BibitemOpen
  \bibfield  {author} {\bibinfo {author} {\bibfnamefont {S.}~\bibnamefont
  {Dwyer}}, \bibinfo {author} {\bibfnamefont {L.}~\bibnamefont {Barsotti}},
  \bibinfo {author} {\bibfnamefont {S.~S.~Y.}\ \bibnamefont {Chua}}, \bibinfo
  {author} {\bibfnamefont {M.}~\bibnamefont {Evans}}, \bibinfo {author}
  {\bibfnamefont {M.}~\bibnamefont {Factourovich}}, \bibinfo {author}
  {\bibfnamefont {D.}~\bibnamefont {Gustafson}}, \emph {et~al.},\ }\bibfield
  {title} {\bibinfo {title} {Squeezed quadrature fluctuations in a
  gravitational wave detector using squeezed light},\ }\href
  {https://doi.org/10.1364/oe.21.019047} {\bibfield  {journal} {\bibinfo
  {journal} {Optics Express}\ }\textbf {\bibinfo {volume} {21}},\ \bibinfo
  {pages} {19047} (\bibinfo {year} {2013})}\BibitemShut {NoStop}%
\bibitem [{Note1()}]{Note1}%
  \BibitemOpen
  \bibinfo {note} {The actual optical loss in the filter cavity will depend on
  the physical realization of $G$, which is beyond the scope of this paper.
  Here we choose $\Lambda _f=0.2\%$ as a reference value, to show that neither
  SNR enhancement nor the stability of the system is compromised in the
  presence of such an effective loss in the filter cavity.}\BibitemShut {Stop}%
\bibitem [{\citenamefont {Vollmer}\ \emph {et~al.}(2014)\citenamefont
  {Vollmer}, \citenamefont {Baune}, \citenamefont {Samblowski}, \citenamefont
  {Eberle}, \citenamefont {Handchen}, \citenamefont {Fiur{\'{a}}{\v{s}}ek},\
  and\ \citenamefont {Schnabel}}]{Vollmer2014}%
  \BibitemOpen
  \bibfield  {author} {\bibinfo {author} {\bibfnamefont {C.~E.}\ \bibnamefont
  {Vollmer}}, \bibinfo {author} {\bibfnamefont {C.}~\bibnamefont {Baune}},
  \bibinfo {author} {\bibfnamefont {A.}~\bibnamefont {Samblowski}}, \bibinfo
  {author} {\bibfnamefont {T.}~\bibnamefont {Eberle}}, \bibinfo {author}
  {\bibfnamefont {V.}~\bibnamefont {Handchen}}, \bibinfo {author}
  {\bibfnamefont {J.}~\bibnamefont {Fiur{\'{a}}{\v{s}}ek}},\ and\ \bibinfo
  {author} {\bibfnamefont {R.}~\bibnamefont {Schnabel}},\ }\bibfield  {title}
  {\bibinfo {title} {Quantum up-conversion of squeezed vacuum states from 1550
  to 532~nm},\ }\href {https://doi.org/10.1103/physrevlett.112.073602}
  {\bibfield  {journal} {\bibinfo  {journal} {Physical Review Letters}\
  }\textbf {\bibinfo {volume} {112}},\ \bibinfo {pages} {073602} (\bibinfo
  {year} {2014})}\BibitemShut {NoStop}%
\bibitem [{\citenamefont {Bentley}\ \emph {et~al.}(2021)\citenamefont
  {Bentley}, \citenamefont {Nurdin}, \citenamefont {Chen},\ and\ \citenamefont
  {Miao}}]{Bentley2021}%
  \BibitemOpen
  \bibfield  {author} {\bibinfo {author} {\bibfnamefont {J.}~\bibnamefont
  {Bentley}}, \bibinfo {author} {\bibfnamefont {H.}~\bibnamefont {Nurdin}},
  \bibinfo {author} {\bibfnamefont {Y.}~\bibnamefont {Chen}},\ and\ \bibinfo
  {author} {\bibfnamefont {H.}~\bibnamefont {Miao}},\ }\bibfield  {title}
  {\bibinfo {title} {Direct approach to realizing quantum filters for
  high-precision measurements},\ }\href
  {https://doi.org/10.1103/physreva.103.013707} {\bibfield  {journal} {\bibinfo
   {journal} {Physical Review A}\ }\textbf {\bibinfo {volume} {103}},\ \bibinfo
  {pages} {013707} (\bibinfo {year} {2021})}\BibitemShut {NoStop}%
\bibitem [{\citenamefont {Dietrich}\ \emph {et~al.}(2016)\citenamefont
  {Dietrich}, \citenamefont {Fiore}, \citenamefont {Thompson}, \citenamefont
  {Kamp},\ and\ \citenamefont {Höfling}}]{Dietrich2016}%
  \BibitemOpen
  \bibfield  {author} {\bibinfo {author} {\bibfnamefont {C.~P.}\ \bibnamefont
  {Dietrich}}, \bibinfo {author} {\bibfnamefont {A.}~\bibnamefont {Fiore}},
  \bibinfo {author} {\bibfnamefont {M.~G.}\ \bibnamefont {Thompson}}, \bibinfo
  {author} {\bibfnamefont {M.}~\bibnamefont {Kamp}},\ and\ \bibinfo {author}
  {\bibfnamefont {S.}~\bibnamefont {Höfling}},\ }\bibfield  {title} {\bibinfo
  {title} {{GaAs integrated quantum photonics: Towards compact and
  multi-functional quantum photonic integrated circuits}},\ }\href
  {https://doi.org/10.1002/lpor.201500321} {\bibfield  {journal} {\bibinfo
  {journal} {Laser {\&} Photonics Reviews}\ }\textbf {\bibinfo {volume} {10}},\
  \bibinfo {pages} {870} (\bibinfo {year} {2016})}\BibitemShut {NoStop}%
\bibitem [{\citenamefont {Elshaari}\ \emph {et~al.}(2020)\citenamefont
  {Elshaari}, \citenamefont {Pernice}, \citenamefont {Srinivasan},
  \citenamefont {Benson},\ and\ \citenamefont {Zwiller}}]{Elshaari2020}%
  \BibitemOpen
  \bibfield  {author} {\bibinfo {author} {\bibfnamefont {A.~W.}\ \bibnamefont
  {Elshaari}}, \bibinfo {author} {\bibfnamefont {W.}~\bibnamefont {Pernice}},
  \bibinfo {author} {\bibfnamefont {K.}~\bibnamefont {Srinivasan}}, \bibinfo
  {author} {\bibfnamefont {O.}~\bibnamefont {Benson}},\ and\ \bibinfo {author}
  {\bibfnamefont {V.}~\bibnamefont {Zwiller}},\ }\bibfield  {title} {\bibinfo
  {title} {Hybrid integrated quantum photonic circuits},\ }\href
  {https://doi.org/10.1038/s41566-020-0609-x} {\bibfield  {journal} {\bibinfo
  {journal} {Nature Photonics}\ }\textbf {\bibinfo {volume} {14}},\ \bibinfo
  {pages} {285} (\bibinfo {year} {2020})}\BibitemShut {NoStop}%
\bibitem [{Note2()}]{Note2}%
  \BibitemOpen
  \bibinfo {note} {Symmetric logarithmic scaling is logarithmic in both the
  positive and negative directions from the origin except for a narrow range
  around zero in which the plot is linear.}\BibitemShut {Stop}%
\bibitem [{Note3()}]{Note3}%
  \BibitemOpen
  \bibinfo {note} {\protect \url
  {https://github.com/artemiy-dmitriev/vectfit}}\BibitemShut {NoStop}%
\end{thebibliography}%

\end{document}